\documentclass[10pt, pdflatex,twocolumn,iicol,sn-mathphys-num]{sn-jnl}

\usepackage{graphicx}%
\usepackage{multirow}%
\usepackage{amsmath,amssymb,amsfonts}%
\usepackage{amsthm}%
\usepackage{mathrsfs}%
\usepackage[title]{appendix}%
\usepackage{xcolor}%
\usepackage{textcomp}%
\usepackage{manyfoot}%
\usepackage{booktabs}%
\usepackage{algorithm}%
\usepackage{algorithmicx}%
\usepackage{algpseudocode}%
\usepackage{listings}%
\usepackage{tabularx,ragged2e}

\usepackage[utf8]{inputenc}
\usepackage{xcolor}
\usepackage{multicol}
\usepackage{hyperref}
\usepackage{float}
\usepackage{amsmath}          
\usepackage{amssymb}          
\usepackage{amsthm}
\usepackage{float}
\usepackage{graphicx}
\usepackage{parskip}
\usepackage{url}
\usepackage{siunitx}
\usepackage{mathtools}
\usepackage{comment}
\usepackage{enumitem}
\usepackage{pdflscape}
\usepackage{caption, booktabs}
\usepackage{balance}
\usepackage{scalerel}
\usepackage{tikz}
\usepackage{longtable}
\usepackage{multirow}
\usepackage{tabularx}
\usetikzlibrary{svg.path}
\definecolor{orcidlogocol}{HTML}{A6CE39}
\tikzset{
  orcidlogo/.pic={
    \fill[orcidlogocol] svg{M256,128c0,70.7-57.3,128-128,128C57.3,256,0,198.7,0,128C0,57.3,57.3,0,128,0C198.7,0,256,57.3,256,128z};
    \fill[white] svg{M86.3,186.2H70.9V79.1h15.4v48.4V186.2z}
                 svg{M108.9,79.1h41.6c39.6,0,57,28.3,57,53.6c0,27.5-21.5,53.6-56.8,53.6h-41.8V79.1z M124.3,172.4h24.5c34.9,0,42.9-26.5,42.9-39.7c0-21.5-13.7-39.7-43.7-39.7h-23.7V172.4z}                 svg{M88.7,56.8c0,5.5-4.5,10.1-10.1,10.1c-5.6,0-10.1-4.6-10.1-10.1c0-5.6,4.5-10.1,10.1-10.1C84.2,46.7,88.7,51.3,88.7,56.8z};
  }
}
\newcommand\orcidicon[1]{\href{https://orcid.org/#1}{\mbox{\scalerel*{
\begin{tikzpicture}[yscale=-1,transform shape]
\pic{orcidlogo};
\end{tikzpicture}
}{|}}}}

\usepackage{svg}
\usepackage{subcaption}

\begin{document}

\title{Cross-Technology Interference: Detection, Avoidance, and Coexistence Mechanisms in the ISM Bands}

\author*[1]{\fnm{Zegeye Mekasha} \sur{Kidane\orcidicon{0009-0009-5141-8160}}}\email{zkidane@mpifr.de}

\author*[2]{\fnm{Waltenegus} \sur{Dargie\orcidicon{0000-0002-7911-8081}}}\email{waltenegus.dargie@tu-dresden.de}
\equalcont{The authors contributed equally to this work.}

\affil[1]{\orgdiv{Electronics Division}, \orgname{Max Planck Institute for Radio Astronomy}, \city{Bad Muenstereifel}, \postcode{53902}, \state{Rheinland-Pfalz}, \country{Germany}}

\affil[2]{\orgdiv{Faculty of Computer Science}, \orgname{Technische Universit{\"a}t Dresden}, \city{Dresden}, \postcode{01062}, \state{Sachsen}, \country{Germany}}

\maketitle

\abstract[\textbf{Abstract}\\
{
 A large number of heterogeneous wireless networks share the unlicensed spectrum designated as the ISM (Industry, Scientific, and Medicine) radio band. These networks do not adhere to a common medium access rule and differ in their specifications considerably. As a result, when concurrently active, they cause cross-technology interference (CTI) on each other. The effect of this interference is not reciprocal, the networks using high transmission power and advanced transmission schemes often causing disproportionate disruptions to those with modest communication and computation resources. CTI corrupts packets, incurs packet retransmission cost, introduces end-to-end latency and jitter, and make networks unpredictable. The purpose of this paper is to closely examine its  impact on low-power networks which are based on the IEEE 802.15.4 standard. It discusses latest developments on CTI detection, coexistence and avoidance mechanisms as well on messaging schemes which attempt to enable heterogeneous networks directly communicate with one another to coordinate packet transmission and channel assignment.
 }

\keywords[\textbf{Keywords} {Coexistence, Cross-technology interference, heterogeneous networks, Internet of Things, wireless sensor networks}


\section{Introduction}
\label{sec:intro}

The vision of the Internet of Things \cite{sundmaeker2010vision} presupposes the coexistence of and close collaboration between multiple technologies \cite{dargie2010fundamentals}. In precision agriculture, for example, wireless sensor networks and Unmanned Aerial Vehicles (UAV) can be jointly deployed to achieve highly precise sensing and efficient micro resource management \cite{primicerio2012flexible, boursianis2022internet}. The nodes on the ground can collect various soil parameters whilst the UAVs assist in collecting and aggregating sensed data as well as in micro-administering  resources such as herbicide. Similarly, in water quality monitoring, ground nodes can collect such parameters as pH, water temperature, turbidity, and dissolved oxygen, whilst Unmanned Surface Vehicles (USV) aggregate data from these sensors and assist in connecting disconnected regions \cite{dunbabin2009autonomous}. In smart industries, where a seamless interaction between  equipment, robots, and other objects may be required, the integration of multiple technologies is essential to achieve efficient, safe, and reliable operation \cite{sisinni2018industrial, dargie2023monitoring}. 

For multiple technologies to coexist, issues pertaining to medium access and interference have to be resolved first. When the technologies are developed independently but share the same spectrum, they may cause interference on each other. This type of interference is called cross-technology interference, or CTI, in short \cite{hithnawi2014understanding, gollakota2011clearing}. CTI becomes formidable when there is a significant disparity in the bandwidth and power requirements of the technologies, often those relying on low-bandwidth and low transmission power becoming victims of intense CTI. The focus of this paper is to closely examine the problem of CTI emanating from technologies sharing the ISM radio band, a license-free radio band internationally reserved for Industry, Scientific, and Medical purposes \cite{tuch1993development}. The band occupies several ranges in the radio frequency spectrum, but the one which is of particular interest to us is the range between 2.4 and 2.5 GHz with the centre frequency located at 2.45 GHz. This band has a 100 MHz bandwidth and is often employed by short-range, low-power wireless technologies. 

\begin{figure}[t!]
	\centering
	\includegraphics[width=0.5\textwidth]{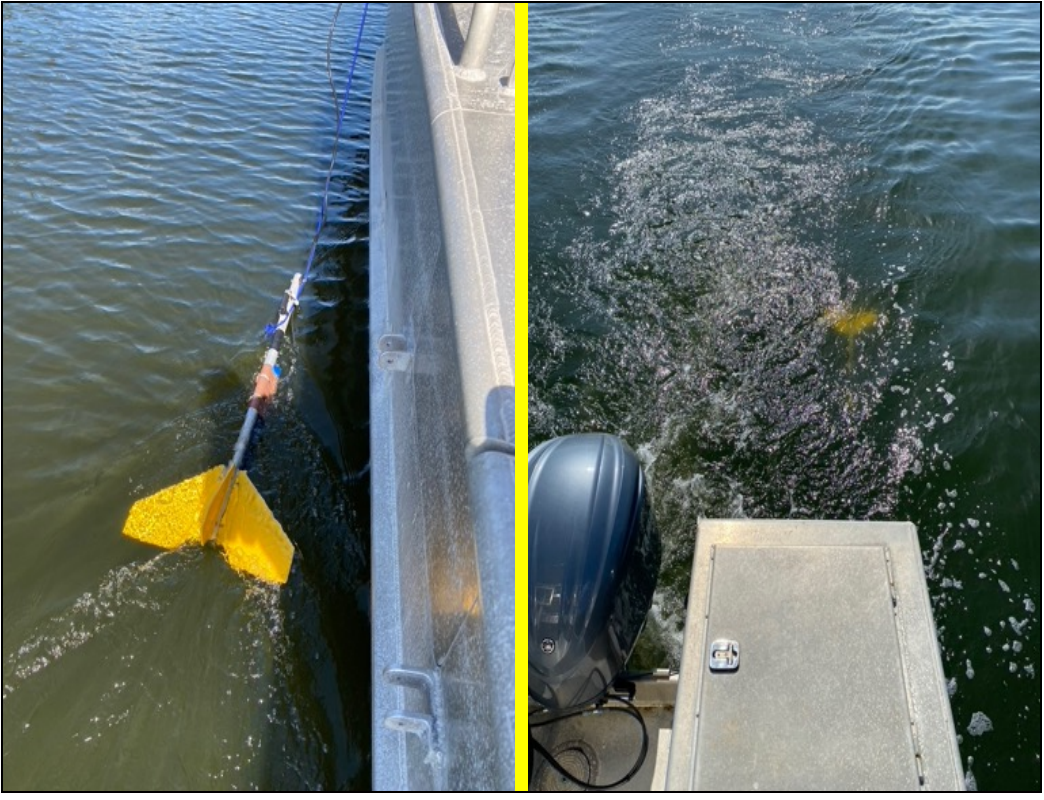}
	\caption{Water quality monitoring at North Biscayne Bay, Miami, Florida. }
	\label{fig:wqm}
\end{figure}

In order to motivate the subject matters discussed in this paper, we begin by relating our experience with CTI. In the summer of 2023  we undertook multiple research expeditions with the Institute of Environment at Florida International University (FIU) on North Biscayne Bay, Miami, Florida. Twice a month, and whenever the need arises, the institute undertakes a boat tour on the bay and its surrounding areas to collect water quality parameters (pH, temperature, conductivity, dissolved oxygen, turbidity, chlorophyll, and fluorescent dissolved organic matter). A tour requires a certified captain and at least one researcher, and lasts about two hours. Fig. ~\ref{fig:wqm} displays the deployment of the  water quality monitoring device. Recently, the quality of the water in the bay has been considerably affected by both natural and man-made causes, giving rise to the death of a substantial amount of fish and other aquatic species \cite{carbonell2021seagrass, bradbury2023site}. In order to achieve a more efficient and scalable monitoring, the institute has started deploying special autonomous (unmanned) surface vessels (USV). Our expedition was intended to experimentally investigate the extent to which wireless sensor networks could be employed to monitor the water quality at a much higher spatio-temporal and scalable resolution. In this regard, an essential requirement was to establish resilient and reliable networks which operate in the presence of a rough water and extreme weather condition. Moreover, the sensor networks should be able to interact with the USVs. 

\begin{figure}[t!]
	\centering
	\includegraphics[width=0.5\textwidth]{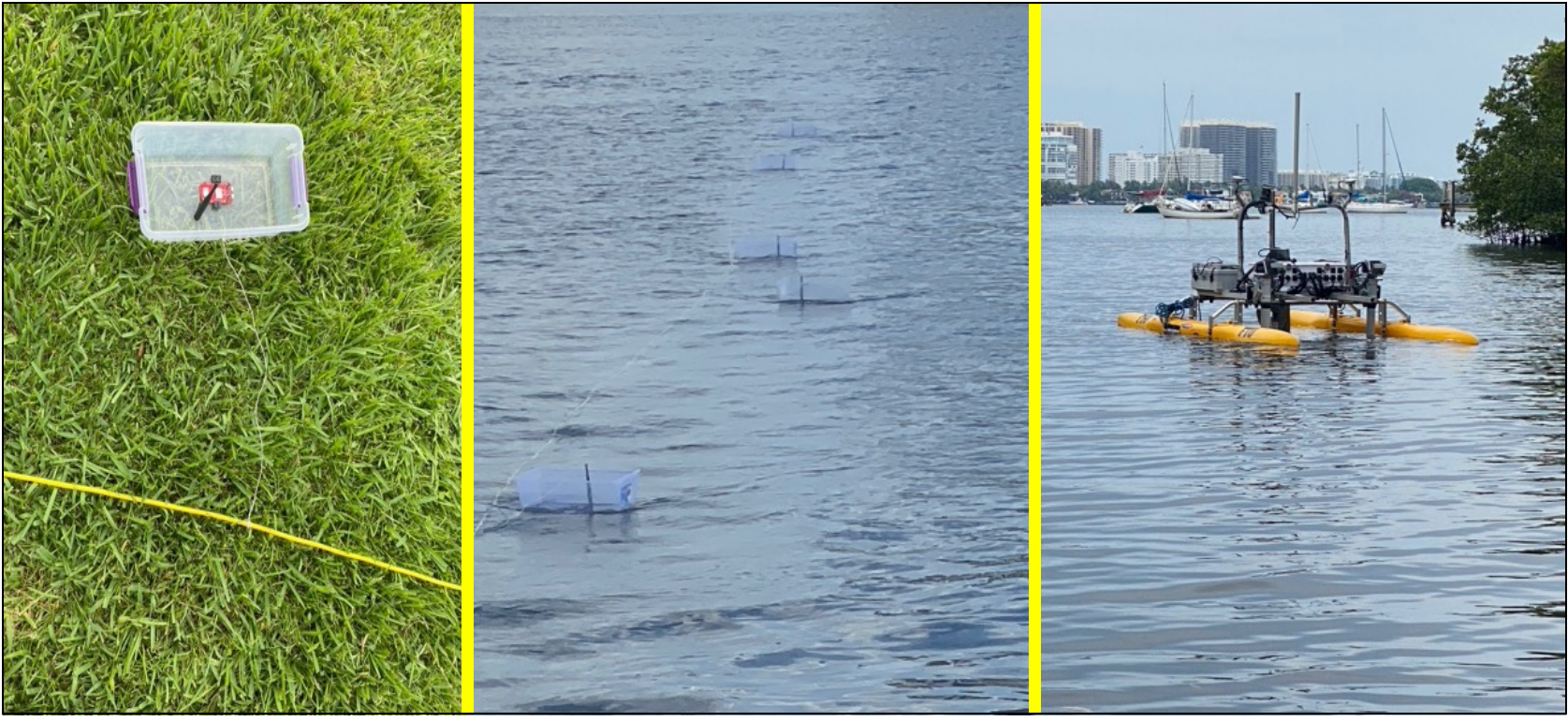}
	\caption{Deployment of an Unmanned Surface Vehicle and a Wireless Sensor Network at North Biscayne Bay, Miami, Florida \cite{dargie2024mitigating}.}
	\label{fig:bbc}
\end{figure}

\begin{figure}[t!]
	\centering
	\includegraphics[width=0.5\textwidth]{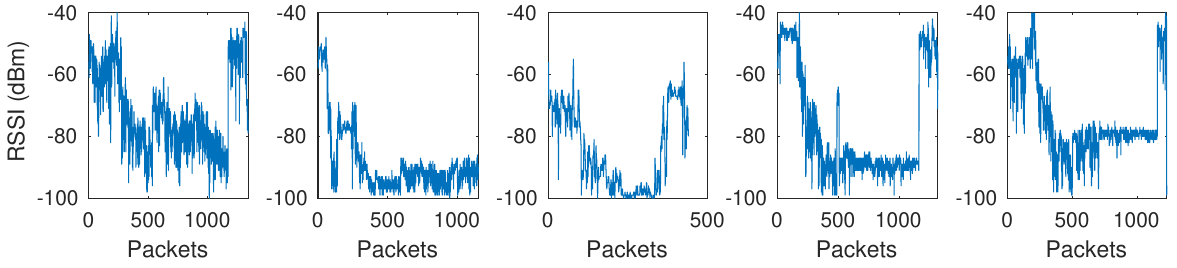}
	\caption{Link quality fluctuation (in terms of the change in RSSI of received packets) in the absence of any CTI. }
	\label{fig:rssi_bbc}
\end{figure}

\begin{figure}[t!]
	\centering
	\includegraphics[width=0.5\textwidth]{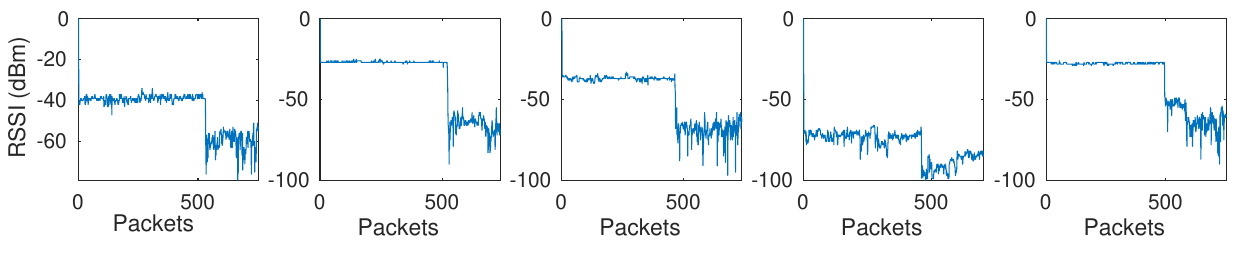}
	\caption{Link quality fluctuation in the absence and presence of a CTI.}
	\label{fig:rssi_bbc_land}
\end{figure}

In one of these expeditions, we deployed a network of six wireless sensor nodes placed in open plastic boxes. The boxes were tied to a long rope which, in turn, was tied to a boat (Fig.~\ref{fig:bbc}). The nodes self-organised using the RPL protocol \cite{clausen2011critical} and a 2.4 GHz radio to support distributed sensing and multi-hop communication. The distance between the nodes was about 50 m. In the absence of CTI, the nodes communicated with one another with modest packet loss, despite an appreciable water motion. Fig. \ref{fig:rssi_bbc} shows the change in the RSSI of successfully transmitted packets, correctly mirroring the movement of the water. When, however, one of the USVs from FIU was within 300 m radius or so, communication was considerably inhibited. We tested all the available channels (there were 16 available non-overlapping channels) to minimise the effect of CTI, but the network performance remained poor. In order to separate the effect of the movement of the water on the link quality from the effect of CTI, we deployed the network along the shore of the bay and observed the link quality fluctuation both in the absence and presence of the USV. As can be seen in Fig. ~\ref{fig:rssi_bbc_land}, the link quality was relatively stable until the USV entered into the interference zone of the network; afterwards, the network transited into an unstable state, followed by a complete disconnection of all the links. The USV was a product of SeaRobotics Corporation\footnote{https://www.searobotics.com/} and employed the IEEE 802.11b standard to interact with its remote control station.      

\begin{figure}[t!]
	\centering
	\includegraphics[width=0.5\textwidth]{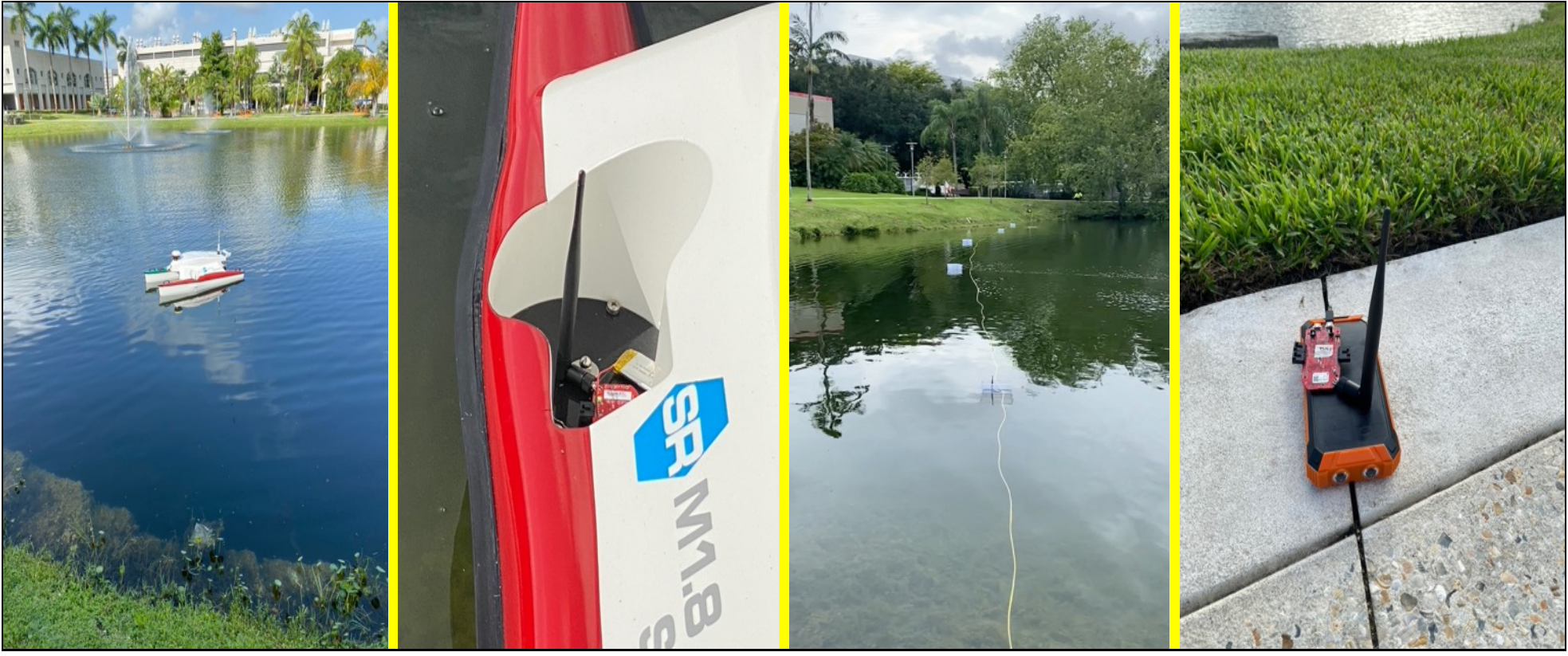}
	\caption{Deployment of an Unmanned Surface Vehicle and a Wireless Sensor Network at one of the Lakes on Florida International University main campus. }
	\label{fig:mmu}
\end{figure}

\begin{figure}[t!]
	\centering
	\includegraphics[width=0.5\textwidth]{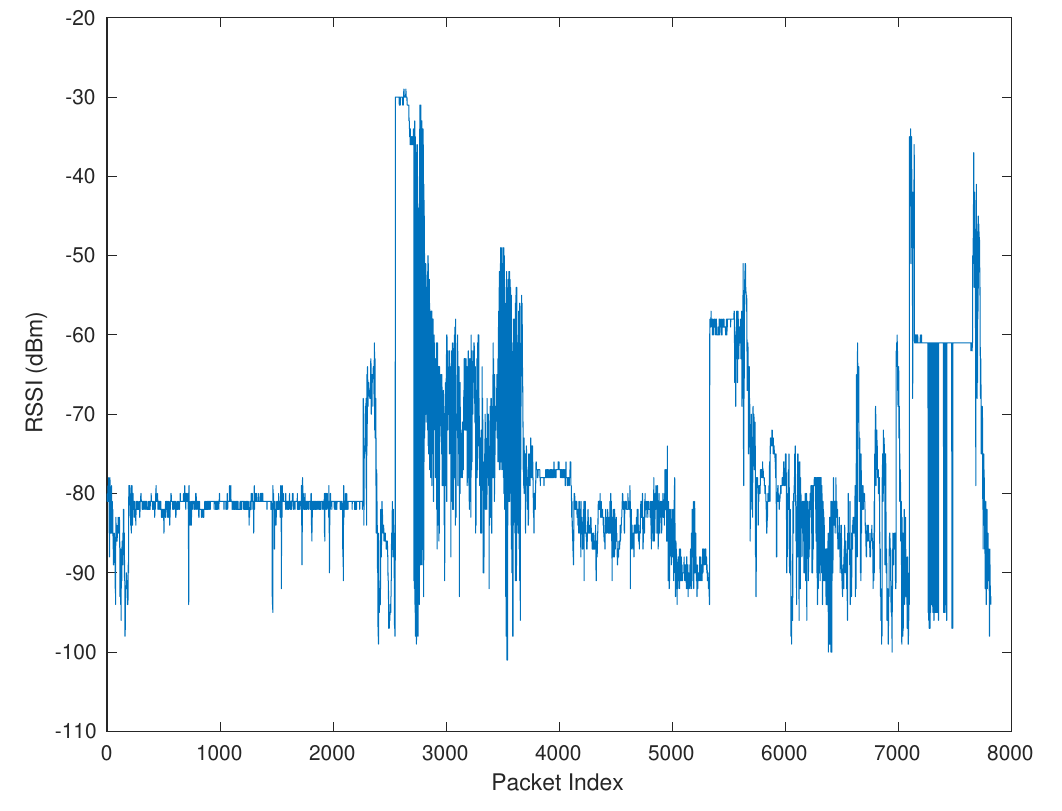}
	\caption{Link quality fluctuation in the presence of CTI (Deployment on a lake on FIU's Main Campus).}
	\label{fig:cti_fiu}
\end{figure}

Our second experience was on one of the lakes on FIU's Main Campus. This time, we collaborated with the team of the Motion, Robotics, and Automation Lab to jointly deploy a wireless sensor network and a USV on the lake. Additionally, one sensor node was deployed on the boat itself. Both this node and the nodes deployed on the surface of the lake communicated with a node placed outside the lake (ref. to Fig.~\ref{fig:mmu}). The present USV had a more complex setup than the one deployed on North Biscayne Bay. It communicated with the remote control station using a proprietary and powerful transceiver, operating in the 4.9-5.8 GHz band, but in addition, the control station was remotely controlled by a human agent using the IEEE 802.11b standard. When the USV navigated autonomously, both the node deployed on the boat and on the lake experienced no interference and the link quality was, by and large, stable; as soon as a human agent interacted with the boat using the IEEE 802.11b interface, all the nodes experienced a significant CTI. The nodes which was affected the worst was the one carried by the autonomous boat. Fig.~\ref{fig:cti_fiu} shows the link quality of this node, as reflected by the RSSI of the packets it received from  the base station.    

 As the first contribution of this paper, our survey includes latest developments on coexistence and CTI avoidance mechanisms. In particular, we review state-of-the-art dealing with Cross-Technology Communication (CTC), which enables heterogeneous networks to exchange information about communication timing and channel occupation. As the second contribution, our paper focuses on protocols and algorithms involving actual implementation and deployment as opposed to those based on simulation. As the third contribution, our survey provides a more complete picture of CTI, addressing spectrum occupation, modulation, emerging networks, interference detection, coexistence and avoidance mechanisms, as well as the impact of CTI on the performance and energy-consumption of low-power networks. 

The remainder of this paper is organized as follows: Section \ref{sec:similar_work}, provides a brief summary of the body of work which is similar to ours. Section \ref{sec:cti} introduces CTI and discusses competing standards. Section \ref{sec:cti_detection} discusses CTI detection strategies. Sections \ref{sec:cti_avoidance}  and \ref{sec:coexist-ISM} review avoidance and coexistence strategies, respectively. Section \ref{sec:system} presents medium access and system support strategies for low-power (IoT) networks dealing with CTI. In Section \ref{sec:impacts}, the impact of CTI on low-power networks will be discussed. Finally, in Section \ref{sec:conclusion}, concluding remarks will be made and open issues will be highlighted.

\section{Related Work}
\label{sec:similar_work}

In the recent past, multiple survey papers have been published with the focus on cross-technology interference. In \cite{Vikramsurvey}, the authors survey papers which address cross technology interference between ZigBee, WiFi, and Bluetooth technologies. The authors' main focus was on interference avoidance mechanisms. In \cite{He_ctc_2022}, the authors survey papers on cross-technology communication (CTC), which enables heterogeneous technologies to coordinate communication and channel assignment. The authors compare the performance of different approaches in terms of throughput, reliability, hardware modification, and concurrency. By contrast, the present paper provides a more comprehensive understanding of CTI and proposed approaches to deal with it. A comprehensive and well-structured review of the role of machine learning in improving the performance of IEEE 802.11 family networks is presented in \cite{Szott_improving_wifi_ml_2022}. The authors identify four distinct features which can take advantage of latest developments in machine learning: coexistence in core WiFi networks, distributed adaptation in emerging WiFi networks (WiFi-6 and WiFi-7), multi-hop networks, and connectivity management. As far as coexistence is concerned, the authors' focus was limited to the coexistence of WiFi networks with LTE networks. Our paper complements theirs by reviewing papers dealing with the coexistence of low-power (IEEE 802.15.4) networks with WiFi and Bluetooth technologies.

\section{Cross-Technology Interference}
\label{sec:cti}

The first step to mitigate CTI is to understand its causes. This concern has to be approached in two ways. The first is to understand the networks which produce the interference; the second is to understand the underlying communication standards and specifications based on which the networks operate. A decade or so ago, the devices which typically produced and were affected by CTI were low-power, low-range, or low-rate devices and networks, such as cordless telephones, microwave ovens, wireless battery chargers, Bluetooth devices, and local area networks. Though these devices and networks can still be a concern, the most formidable challenges come from emerging autonomous systems whose operation and interaction regions are much wider. These devices typically rely on a high transmission power and large antennas; and employ advanced modulation and medium access techniques to ensure safe and reliable operations. For example,  low-altitude enterprise Unmanned Aerial Vehicles (UAV) produced by DJI\footnote{https://www.dji.com} employ long-range proprietary controllers operating in the 2.4 and 5.8 GHz radio bands and switch between these bands using time-slotted and frequency hopping strategies to deal with CTI; in doing so, they themselves produce a considerable CTI to nearby low-power IoT devices and networks.

\begin{table*}[ht!]
    \centering
    \begin{tabular}{|p{3.0cm}|p{3.0cm}|p{3.0cm}|p{3.0cm}|p{3.0cm}|}
 \hline
 \multicolumn{5}{|c|}{Coexisting wireless device comparison} \\
 \hline
 Technology &IEEE 802.15.4 &IEEE 802.15.1 &IEEE 802.11 b/g/n & Commercial drone (UAV) Technology \\
 \hline
Number of Channels   & 16   &79        &11/13/14 &  Use 802.11 channels \\
\hline
Data rate    & 250 Kbps   & 1 Mbps    &11 Mbps, 54 Mbps &  90-100 Mbps \\
\hline
Band width   &2 MHz    & 15 MHz       & 20 MHz & 80 MHz \\ 
 \hline
 Transmit Power   &0 dBm   & 1-20 dBm       & 10-20 dBm & 10 mW-1 W  \\ 
 \hline
 Transmit range[m]   &1-100    & 1-10       & 1-100 & 100-500   \\
  \hline
  Power consumption   &Very Low    & Low        & High  & Very High   \\
  \hline
   Throughput   &20-250 Kbps    & 720 Kbps       & 11 Mbps & 90 Mbps  \\
  \hline
  Channel allocation   &11-26    & 1-79        & 1-13  &   NA \\
  \hline
  Method of Channel Access   &CSMA/CA DCS, Scheduling  & Master Slave scheme, AFH & CSMA/CA, DCS &  NA \\
  \hline  
  Operating Frequency   & 2.4 GHz   & 2.4 GHz & 2.4 GHz & 2.4 GHz, 5.8 GHz, 433 MHz, 915 MHz   \\
  \hline
  Frequency Modulation   & Orthogonal-PSK, BPSK   & GFSK & OFDM, DSSS,CCK  & BPSK, OFDM, OTFS \\
  \hline    
  Coexistence   & CCA based energy detection   & FH, spread spectrum    & CCA based energy detection & spectrum overlay multiple access (SOMA),TDMA  \\
  \hline   
  Encryption   & AES-BC (CTR)   & E0-SC  & RC4-SC(WEP), AES-BC  & AES, ARIA, HMAC, ECDSA  \\
  \hline 
  Data protection   & 16-bit CRC   & 16-bit CRC   & 32-bit CRC & CRC \\
  \hline   
  Interference Avoidance Method   & Fixed Collision avoidance, Frequency Hopping, TSCH   & Frequency Hopping  & Collision avoidance & Frequency Hopping \\
  \hline
  Signal On-air time & [576, 4256] us & 366 us & [194, 542] us &    [1, 3] minutes \\
  \hline
  Minimum packet interval & 2.8 ms or 192 us & NA & $\geq$ 28 us & 0.38 ms    \\
  \hline
  Peak to average power ratio & $\leq$ 1.3 & $\leq$ 1.3 & $\geq$1.9 &  $\geq$2 \\
  \hline
\end{tabular}
    \caption{Comparison of coexisting ISM frequency band technologies.}
    \label{tab:coexist_compare}
\end{table*}

\subsection{Standards}

Most of the wireless links which potentially cause and are affected by CTI in the ISM radio band are those based on the IEEE 802.11 and IEEE 802.15 standards. More specifically, in the former, we have the IEEE 802.11 b/g/n/ax/be/ba networks (commonly known as WiFi networks), whereas in the latter, we have IEEE 802.15.1 (Bluetooth), 802.15.3 (high-rate PAN), IEEE 802.15.4 (low-rate PAN), and IEEE 802.15.5 (Mesh) networks. The IEEE 802.15.4 standard defines 27 channels, each having a bandwidth of 2 MHz (refer to Table \ref{tab:coexist_compare}). Of these, 16 channels (designated Channels 11-26) are in the 2.4 GHz frequency band; Channels 1 to 10 are in the 915 MHz frequency band; and Channel 0 is in the 868 MHz frequency band. Equation \ref{eq:channel_center_freq} is used to determine  the centre frequencies of Channels 11 to 26. Similarly, Bluetooth has 79 channels, each having a bandwidth of 1 MHz; and IEEE 802.11/b/g/n wireless local area networks have 14 available channels. Of these, only 11 are legally available in the US and only 13 are available in Europe. Nevertheless, most existing networks use only 3 of these (Channels 1, 6 and 11) to avoid adjacent channel interference. Channel 1 potentially interferes with Channels 11, 12, 13 and 14 of the IEEE 802.15.4 ISM band; Channel 6 potentially interferes with Channels 16, 17, 18, and 19 of the IEEE 802.15.4 ISM band; and Channel 11, overlaps with Channels 21 to 24. All available Bluetooth channels overlap with IEEE 802.15.4 channels, although, as independent studies show, interference from Bluetooth technologies is not appreciable \cite{shin2007packet, Garroppo_expermentCoexistence}.

The spectral assignment of the IEEE 802.15.4 channels is given as:

\begin{equation}
\label{eq:channel_center_freq}
    F_{c}(n) = 2405 + 5 (n-11)   
\end{equation}
where $F_c$ is the centre frequency in MHz and $n= 11,12,... 26$ is the channel number.

\subsection{Link Quality Metrics}

Recent work on interference classification relies mainly on mapping the pattern of some low-level link quality metrics to well-known classes of interference. The pattern itself can be established by sampling the values of these metrics in time and in frequency domains and across all the available channels to establish multi-dimensional distributions and to compare the distributions with the signal profiles of the different standards discussed above. This approach performs poorly when there are multiple sources of interference. More importantly, there is no generally accepted mechanism to map the metrics to any particular physical parameters. Different chip manufacturers often implement their own metrics to characterise link quality.

There are, however, some widely accepted abstract (or coarse-grained) metrics which can be employed to characterise wireless links. The reason we label them as ``coarse-grained`` is that they are obtained by averaging the received power of a signal over a period of time. The IEEE 802.15.4 specification \cite{adams2006introduction} defines two physical layer parameters: Energy Detection (ED) and Link Quality Indicator (LQI). ED is an approximation of the received signal's power within the bandwidth of a given channel. It is obtained by calculating the average power corresponding to 8 successive symbols. Ideally, its range is 40 dB (in some technologies, even 85 dB) within which the mapping from a received power in decibels to an ED value is linear, with an accuracy of $\pm$ 6 dB. The metric is useful for assessing background noise and for calculating LQI. Another widely employed metric is the Received Signal Strength Indicator (RSSI), which is often associated with successfully received packets. In both cases, the source of a received signal may be an interferer. Hence, both metrics may not directly correspond to the quality with which a packet is received. The LQI, on the other hand, characterises the strength and/or quality of a desirable signal. Hence, it is directly associated with the quality with which a packet was successfully received. The specification does not state how this metric should be computed, but suggests that it can be determined based on ED, a signal-to-noise ratio estimation, or a combination of both.

\begin{figure}[ht!]
	\centering
	\includegraphics[width=0.50\textwidth]{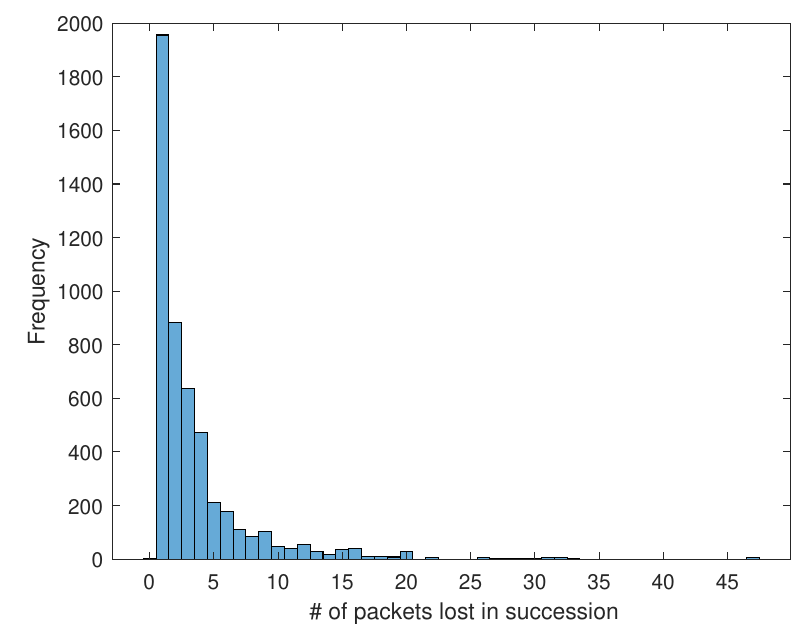}
	\caption{Histogram of a successively lost packet when a node carried by a UAV interacts with a ground wireless sensor node}
	\label{fig:hist_lost_packets}
\end{figure}

Both RSSI and LQI say little about lost packets, so that the exact link state at the time the packets are lost cannot be established. There are some higher-level metrics which attempt to provide an estimation of the link quality of lost packets. One of these is the Packet Reception Rate (PRR). This metric is expressed as the ratio of the number of packets successfully received in the past $\tau $ seconds to the ideal number of packets that could be transmitted in that same time. The metric can be computed using a sliding window. Taking the difference in PRR of adjacent time slots enables to estimate the short-term link stability condition, whether this condition refers to a consistently bad or consistently good state. A CTI can also be indirectly determined by characterising link quality fluctuation in terms of the number of packets successively lost or received \cite{wen2021evaluation}. Assuming that the transmission pattern of the interferer is statistically stable, this metric enables to estimate the average duration a wireless link is under the influence of interference \cite{dargie2021link, dargie2024mitigating}. Fig. \ref{fig:hist_lost_packets} shows the histogram of successively lost packets when a node carried by a UAV transmitted packets in burst to a ground node. As can be seen, the histogram can be modelled as an exponential probability density function having the form: $f_x(x) = \lambda e^{-\lambda x}$, where $x$ is the number of packets lost in succession and $\lambda$ encodes the rate at which the graph decreases. Once $f_x(x)$ is estimated, it is possible to compute the average number of packets lost in succession: $E[x] = \int x. f_x(x)$. With this and the knowledge of packet length and the node's transmission rate, it is possible to determine the average duration of interference. This approach assumes that the main cause of packet loss is CTI.  

\subsection{Wireless Links Models}

Characterising wireless links is a classical problem. In cellular (mobile) networks, for example, models of wireless links are useful for radio management (for adapting transmission power, time-slots appropriation, handover, etc.). In such networks, both transmitters and receivers establish channel state information (CSI) by scanning the available channels at an appreciably high rate; then various features of this information are extracted to determine the characteristics of the wireless links and to carry out some compensatory or adaptive measures. Low-power networks do not have sufficient resources to perform elaborate computation. Consequently, they often rely on offline, light-weight link models to achieve the same goals.

Besides CTI, the quality of a wireless link may be affected by several external factors, including surrounding temperature, heavy rain, humidity, and shadowing \cite{dargie2024estimation}. In order to distinguish a CTI from link quality deterioration arising from factors inherent to the deployment setting, some researchers have proposed models to account for the contribution of the latter (which are thought to be more persistent). For example, in the context of a joint deployment consisting of a UAV and a wireless sensor network, Dragulinescu et al. identified six types of propagation environments \cite{Dragulinescu_USV_IOT}: free space, rural area, suburban, urban, dense urban and highly dense urban, though their study specifically addresses rural and free space. In both cases, the authors further distinguish between aerial and terrestrial channels, the former models the characteristic of a channel established between a UAV and a ground node, whereas the latter models the characteristics of a channel established between two ground nodes. Likewise, an air-to-air channel models the characteristics of a channel established between two or more UAVs or a UAV and a satellite. Habib et al. \cite{Habib_overseaChannel}, Kim et al. \cite{Kim_UAVProbability}, and Dargie et al. \cite{dargie2024link} likewise propose additional models to characterise signal propagation involving different water bodies (sea, river, lake). 

\section{Detection}
\label{sec:cti_detection}

Detecting the causes and the characteristics of a CTI is crucial to effectively deal with it. A spectrum analysis is required to establish a complete information, but this involves, among others, Fast Fourier Transform (FFT), which is costly, both in terms of the resources it demands and the time it takes. All other approaches are at best approximations. One of the most frequently employed approaches consists of energy detection, in which a receiver scans all the available channels, probing the power at its radio front end. Finally, it compares some statistical aspects of the power (such as mean, max, min, average, zero-crossing, etc.) with some existing patterns to determine CTI and its potential sources. This approach requires a fast scan time, an appreciable local memory to save the sampled power and reference profiles, as well as some computation. Hence, its reliability depends on the available resources and the extent to which the reference models represent the underlying reality. 

One way of establishing a reference model for a particular source of a CTI is mapping its interaction pattern to the standard to which it complies. A standard is by definition a set of rules. In all communication standards, the specific steps communicating partners take to exchange packets using a shared medium are well-defined. For instance, in most wireless standards, packet transmission is preceded by the transmission of preambles to enable signal detection and synchronization. The length and transmission duration of the preambles vary from standard to standard. Similarly, when a transmitter occupies a medium, the rules to which it complies can be inferred from its manner of packet transmission. For instance, in IEEE 801.11a an idle duration of 16 us (inter-frame space) will be experienced between the transmission of a data packet and the reception of an ACK packet; whereas this duration is 10 us in IEEE 802.11n (2.4 GHz) networks. Similar approaches map energy distributions to known traffic patterns. For example, Qin et al. \cite{qin2020enhancing} observe that WiFi frames are highly clustered and there are small idle leaks within the frame clusters and large white space between frame clusters.  

There are other approaches which rely on low-level features, but they presuppose knowledge of the data encoding and modulation of corrupted packets. In IEEE 802.15.4, the data to be transmitted is first grouped into 4-bit symbols, each of which is then converted into a predefined 32-bit long pseudo-random noise sequence \cite{hithnawi2016crosszig}. The bits (so called chips) are further grouped into even and odd chips and are modulated using Offset Quadrature Phase-Shift Keying (O-QPSK) -- the even chips modulated as In-phase (I) component of the carrier, and the odd chips, as Quadrature (Q) component of the carrier. There is a time offset between the Q-phase chips and the I-phase chips, so as to enable a continuous phase change. Since the message is encoded in the pattern of the phase-shift, the wave shape and amplitude of the carrier is intact. This enables the reliable detection of the modulated signal in the presence of appreciable noise. During packet reception, demodulation and low-level source decoding take place in the first place, inside the radio chip. In most cases bit-by-bit analysis of received packets is not possible, as this is done by a hardware component. Therefore, the lowest-level information available is in the form of 4-bit symbols \cite{hermans2013sonic}.

\begin{figure*}[t!]
	\centering
	\includegraphics[width=0.9\textwidth]{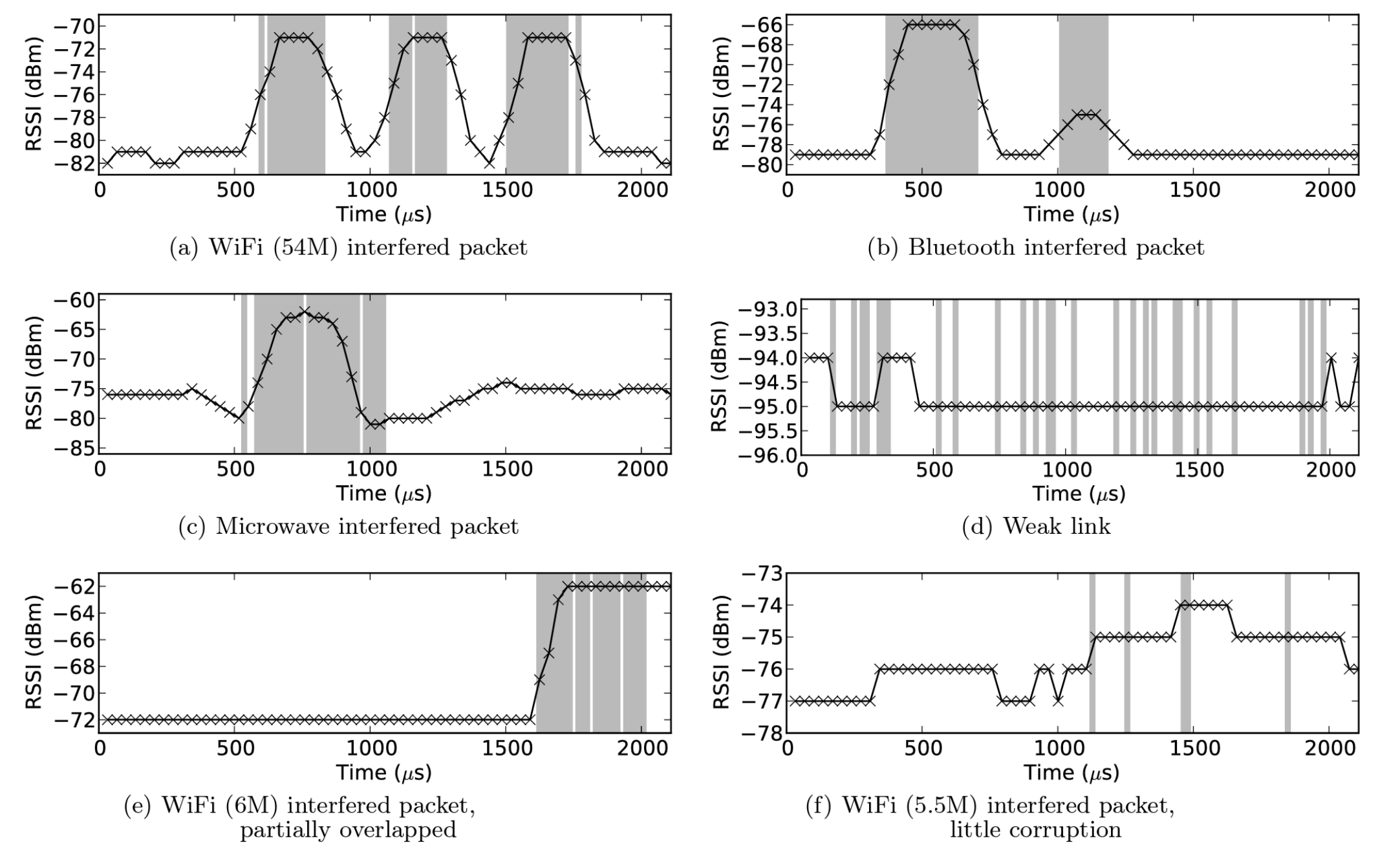}
	\caption{Determining the sources of cross-technology interference by establishing the pattern of corrupted symbols in received packets \cite{hermans2013sonic}.}
	\label{fig:pattern}
\end{figure*}

In \cite{hermans2013sonic} and \cite{hithnawi2014understanding}, the authors closely investigate the symbols of corrupted packets to discover distinct interference patterns which can be traced back to specific causes. The authors distinguish between lost packets and corrupted packets. In the first, a receiver fails to detect packet preambles and subsequent headers which enables it to successfully decode the packet; whereas in the second, a packet is actually received but its CRC flags an error. In \cite{hermans2013sonic}, when a receiver receives a corrupted packet, it requests a retransmission and carries out a symbol-by-symbol comparison to localise the corrupted symbols and establish a pattern. The authors observe that different interference sources leave distinct patterns. Interestingly, corrupted symbols are likely to be received with a relatively high energy (i.e., high RSSI values). The authors trained a supervised model with these and additional low-level features to classify different sources of a CTI. Fig.~\ref{fig:pattern} shows the CTI patterns of different causes based on the analysis of corrupted symbols of received packets. Similarly, Hithnawi et al. \cite{hithnawi2014understanding} purport that interference errors occur in bursts and can be localized to short intervals. By contrast, corrupted bits due to random channel variation are randomly scattered. To characterise error burstiness due to CTI, the authors artificially induced interference using different technologies and analysed the traces of corrupted symbols, counting the frequency of symbol error bursts of various length, for each interferer technology; variable packet length, power level, and distance. 

Similarly, in \cite{YangTamingCTI_For_WiFiZigBee}, the interference patterns of WiFi networks on IEEE 802.15.4 networks is studied experimentally. In the presence of 20 active WiFi APs deployed in the basement of a school building, two sensor nodes exchanged packets every 30 ms at a transmission rate of 250 kbps and using a variable transmission power, ranging from -65 to -27 dBm. The packet size was 127 bytes. To compensate for the effect of CTI, the transmitter's power was adjusted every 10 seconds, from the minimum to the maximum level. Likewise, the transmitter and the receiver simultaneously switched to a different channel every 600 s, beginning from Channel 11 to Channel 25. The entire experiment lasted 8 hours. The experiment results reveal that the CTI patterns of the channels overlapping with Channel 1, 6, 11 of the WiFi network had similar patterns: The duration and interval between corrupted symbols suggest that CTI having a short duration but occurring in short intervals was experienced more frequently than CTI having a long duration or occurring in long intervals. The authors attribute this to short and frequent WiFi data transmissions.

\begin{table*}[h!]
\centering
\begin{tabular}{ |p{3.2cm}|p{3.8cm}|p{3.8cm}|p{3.8cm}|  }
 \hline
 \multicolumn{4}{|c|}{CTI Detection} \\
 \hline
 Research Papers & Research aim & Detection method  &Technical features\\
 \hline
 \cite{kotsiou2019blacklisting, gomes_MABO-TSCH, zorbas2018local, IyerDetectingAvoiding}  &Identification of bad radio channels  & Blacklisting channels with low PDR and RSSI burst   & Establishing reliable wireless links \\
  \hline
 \cite{hermans2013sonic, hithnawi2014understanding, YangTamingCTI_For_WiFiZigBee} & Interference classification & Symbol-level correlation of corrupted packets using correlation and soft values & Mitigation of short-duration WiFi interference in long-duration ZigBee communications\\
  \hline
 \cite{Pulkkinen_cti_detection_2020, nguyen_rfdeeplearning_2024, croce_detect_cti_2018} &Signal modelling and interference detection & Using deep learning and transfer learning for multi-channel spectral representation & RF emission detection, classification, and spectro-temporal localization\\ 
 \hline
 \cite{Uy_design_interdetector}  & Channel state analysis, interference characterisation/classification & Interference detection; ranking of the relative strength of interference & Interference duration estimation\\
 \hline
 \cite{Mathony_interDetction} &Interference and intrusion detection& Use of Random Forest Machine Learning to classify interference & Analyse of received In-phase (I) and Quadrature-phase (Q) samples \\
 \hline
\end{tabular}
\caption{Review of CTI detection mechanisms}
\label{table:detection}
\end{table*}

Croce et al. \cite{croce_detect_cti_2018} employ an Artificial Neural Network (ANN) and a Hidden Markov Model (HMM) to detect CTI during the reception of a WiFi frame. The models enable to differentiate between different sources of interference (IEEE 802.15.4, LTE, microwave, and IEEE 802.11). The authors argue that, whereas for WiFi standard frames the error probability varies during frame reception in different frame fields (PHY, MAC headers, and payloads), errors due to CTI appear randomly when the demodulator attempts to receive exogenous interfering signals. Based on their probability of occurrence, patterns (sequence of occurrence), and time intervals, the errors are classified into different sources. The error analysis in a WiFi receiver has been carried out by demodulating a sequence of completely random bits. The demodulated bits are interpreted  according to the format of a WiFi frame. The authors trained and tested their models using an of-the-shelf WiFi network interface card (NIC) and two types of IEEE 802.15.4-compliant transceivers (CC2420\footnote{https://www.ti.com/product/CC2420. Last visited: 01 October 2024, 11:42 AM CET.} and MRF24J40\footnote{https://ww1.microchip.com/downloads/en/DeviceDoc/39776C.pdf. Last visited: 01 October 2024, 11:44 Am CET.}). The later produced controlled interference, a single source producing interference on the  WiFi link at Channels 11, 10 or 8 with different interference transmitter power levels at a time. Two different approaches were used for classifying  the interference sources acting on the targeted WiFi receiver. The first approach was based on the receiver behaviour in a fixed time interval (a few tens of $ms$) corresponding to a few samples of the error vectors. The second approach was based on a given error burst delimited by the channel busy register (a single frame radiation period). In this case, idle times between consecutive error burst were not considered for classification. The authors claim that with 95\% accuracy they were able to determine the source and timing of interference.
            \begin{figure}[ht!]
            	\centering
            	\includegraphics[width=0.50\textwidth]{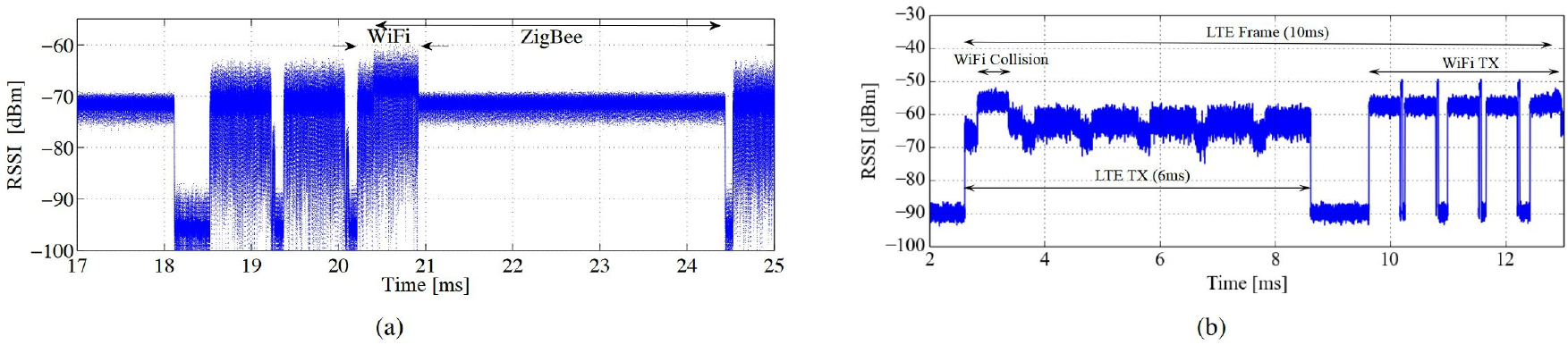}
            	\caption{Interference between technologies: (a) RSSI  detected WiFi-ZigBee Power Signal  and (b) WiFi-LTE collision \cite{croce_detect_cti_2018}}
            	\label{fig:wifi_zigbee_lti_collision}
                \end{figure}

Nguyen  et al. \cite{nguyen_rfdeeplearning_2024} propose a wide band, real time, Spectro-Temporal RF Identification system to detect, classify, and locate Radio Frequency (RF) emissions in time and frequency using RF samples of 100 MHz bandwidth spectrum. The systems combines a one-stage object detection deep learning network with the YOLO object detection algorithm \cite{jiang2022review}. Accordingly, wide-band RF samples are transformed into a 2D time-frequency image based on which individual and overlapping RF emissions are identified, localized, and classified. The transformation of the raw RF samples to visual data follows after the I/Q data stream are divided into equal chunks and N-point FFT is applied to each chunk. The authors report a detection accuracy of 99\%. 

            \begin{figure}[ht!]
            	\centering
            	\includegraphics[width=0.50\textwidth]{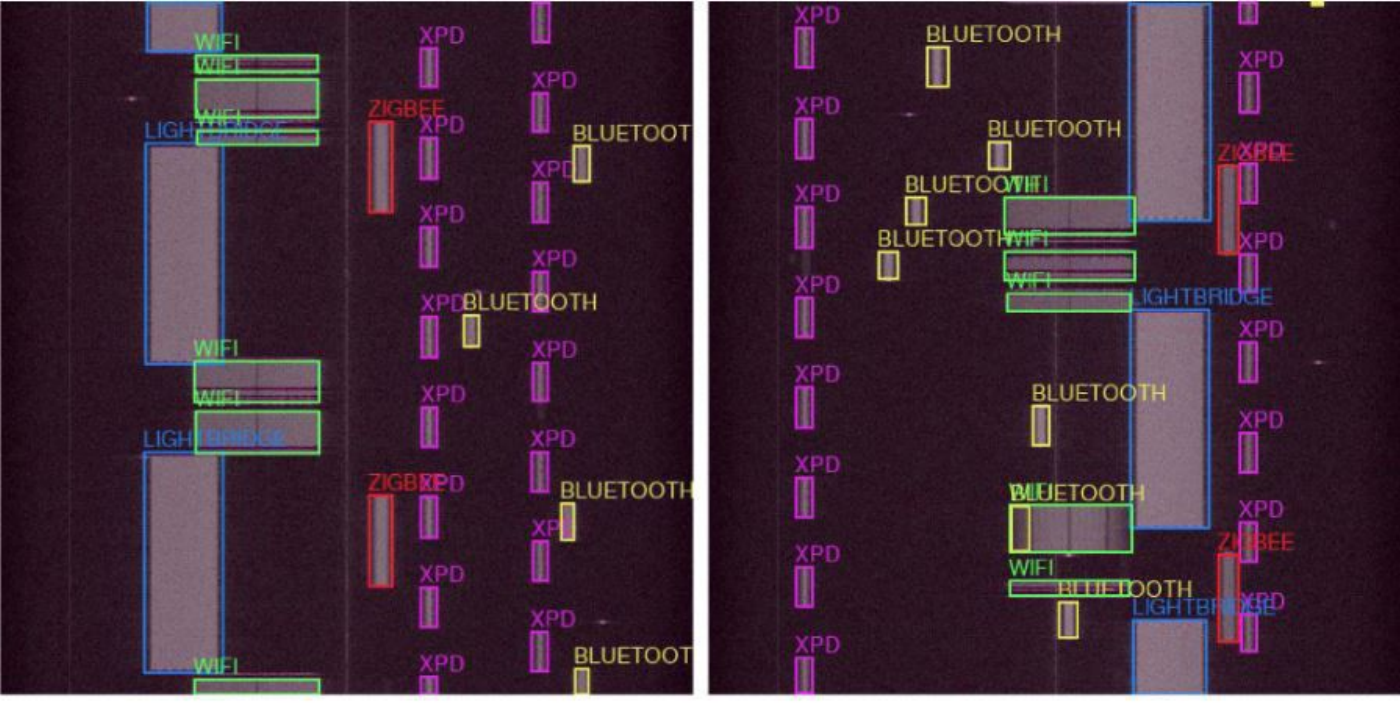}
            	\caption{Different wireless devices RF detection in congested environment\cite{nguyen_rfdeeplearning_2024}}
            	\label{fig:rf_detection}
                \end{figure}      
Table \ref{table:detection} summarises some of the CTI detection approaches we reviewed in this section.

  In \cite{Grimaldi9298759, Grimaldi8570750}, the authors compare different lightweight machine learning classification models, observing that RSSI signal traces with bandwidth in the order of a few kHz lead to a poor representation of the signal envelope and undermine the classification accuracy of single signal bursts. Alternative to RSSI, the authors propose to directly detect the envelop of a received signal. The authors demonstrated that this can be carried out with off-the-shelf IEEE 802.1.4-compatible devices. According to the authors, the spectral features of the envelop reveal rich insights which are useful for training different machine learning models. This approach not only enables the detection and classification of interference sources, but also is able to resolve multiple sources when they cause interference at the same time.  

\section{Avoidance}
\label{sec:cti_avoidance}

Even though avoidance strategies are in essence coexistence strategies, we discuss them separately because they commonly rely on dynamic channel selection. Of course, CTI avoidance can also take place in space and time as well. The former requires the adaptation of some aspects of the physical layer (transmission power, antenna selection, antenna positioning, antenna orientation, etc.) which incurs either a considerable performance penalty to all the networks concerned (but, particularly, to the low-power networks) or restriction on their mobility. The latter, too, requires some form of coordination, the estimation of the optimal length of a time frame and the number of time slots as well as time synchronization. Discussion on these aspects is differed to Section \ref{sec:system}. Dynamic channel selection (frequency hopping) overcomes all these restrictions, but consists of identifying the best channel; determining the timing of channel hopping; and a strategy to convey channel transition between communicating partners. According to Incel et al. \cite{incel2011survey}, proposed approaches supporting dynamic channel selection in IEEE 802.15.4 networks can be classified based on 7 principal aspects. These are: 
\begin{enumerate}
    \item The purpose of channel selection.
    \item The channel assignment strategy.
    \item The presence or absence of a control channel.
    \item Centralised vs. decentralised approach.
    \item Whether all the nodes in the networks communicate using a single channel or whether the use of multiple channels is supported.
    \item The medium access mechanism.
    \item Whether the strategy supports packet broadcasting.
\end{enumerate}

To these aspects we identify one additional feature, namely, whether channel selection is carried out by a transmitter or a receiver. Most existing approaches are transmitter-initiated, distributed, and enable the use of multiple channels. In Section~\ref{sec:system}  we shall discuss in detail the Time-Slotted Channel Hopping (TSCH) protocol -- one of the protocols in low-power networks which are integrated in the IEEE 802.15.4 specification -- and some variants of it. 

Tytgat et al. \cite{tytgat2014analysis} propose a receiver-initiated dynamic channel selection intending to optimise the total average packet error rate (PER) in the network. The authors attempt to achieve this goal -- even though nodes make local decisions -- by defining a metric which aims to identify a channel with the lowest average measured channel power. The authors compare this metric with three competitive metrics, namely,
\begin{itemize}
    \item ``activity'': ranks CTI based on averaged, minimum, and maximum measured channel powers, thus: (avg - min)/(max - min).
    \item  ``min'': selects the channel wherein the minimum measured channel power is the lowest as a receiving channel.
    \item  ``max'': selects the channel wherein the maximum measured power is the lowest at a receiving channel.
.\end{itemize}

Channel ranking is carried out at runtime, with each node sampling the interference levels on the different channels. According to the authors, packet error rate (PER) is minimised when each individual node selects the channel with the estimated least average PER. Nodes wishing to communicate with this node, should discover its receiving channel, in the same way WiFi devises identify the channel at which an access point listens or a Bluetooth slave device following the hopping sequence of its master.

Channel blacklisting \cite{kotsiou2019blacklisting, gomes_MABO-TSCH, zorbas2018local} facilitates channel selection in the presence of persistent CTI. Iyer et al. \cite{IyerDetectingAvoiding} carried out interference detection, classification, and channel recommendation (blacklisting) based on the offline evaluation of various WiFi and Bluetooth traffic channel utilisation characteristics and their impact on low-power IEEE 802.15.4 networks. Interference patterns are established by classifying RSSI bursts using predetermined RSSI intervals, persistent duration, and variance (a clustering component groups together RSSI bursts which are likely to come from the same CTI source). In a subsequent step, the patterns are mapped to different network traffic models using a k-mean classification. This results in distinguishing between different data traffics (WiFi beacons, periodic and non-periodic channel traffic) and in estimating the number of sources transmitting periodic signals (WiFi access points). The authors evaluated their approach using a network of 85 sensor nodes, each node generating 1 packet per minute over a two-hour duration. Interference avoidance has been carried out by ranking the channels according to their RSSI patters. The experiment results suggest that an improvement of a 30\% average throughput can be achieved in comparison to other interference avoidance techniques. This achievement was mainly due to the blacklisting of channels which were likely to suffer from CTI arising from nearby WiFi networks. 

\begin{figure}[ht!]
            	\centering
            	\includegraphics[width=0.50\textwidth]{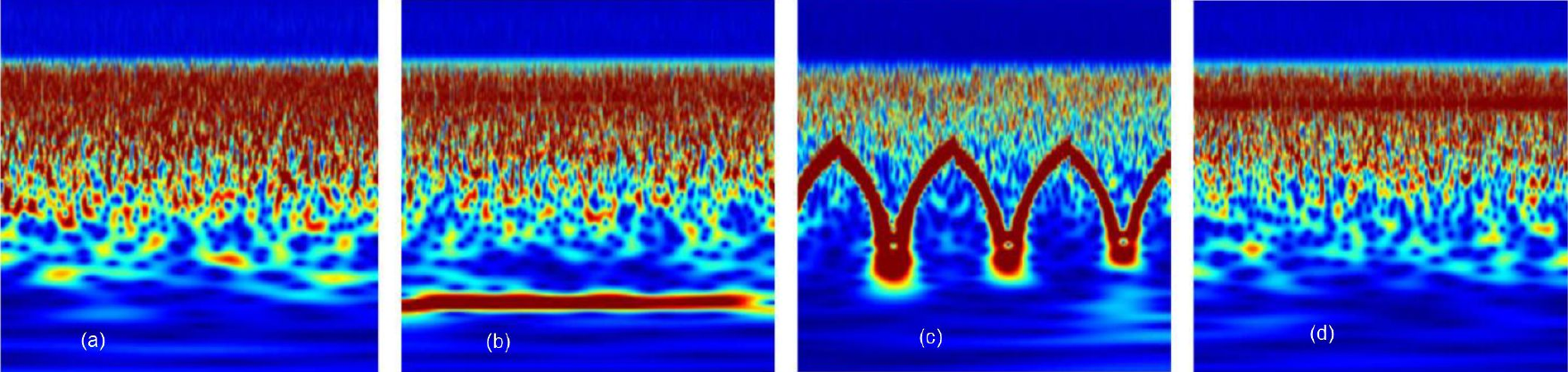}
            	\caption{A scalogram plot of the RFI data set from \cite{Ujan_RFI_recognition_2020}. (a): The original signal of interest with no interference; (b): The signal of interest(SoI) with multi-continuous wave interference; (c): The signal of interest with chirp interference; (d) The signal of interest with continuous wave interference.}
            	\label{fig:scalogram_rfi}
                \end{figure}

More recently, researchers have started to employ machine learning to model and mitigate the impact of CTI. In \cite{Ristea_fCN_radar_rfi_2020}, the authors propose a convolutional neutral network (CNN) to transform a noisy spectrogram into a clean range profile for radar sensors. The model was trained with different real world radar signals (with and without interference and captured by NXP TEF810X 77 GHz radar transceiver \cite{ng2014fully}) representing different range profiles. The authors introduce two metrics to evaluate the performance of their model, namely, the probability of false alarm and the mean absolute error. The former is tested by measuring the receiver operating characteristics curve at various threshold and the later measures the error between the range profile amplitude of targets computed from a level signal and a predicted target signal. In \cite{Mun_automotive_radar_interfere_2020}, a recurrent neutral network (RNN) is proposed to mitigate interference between Frequency Modulated Continuous Wave (FMCW) \cite{wang2014application} and OFDM radar signals. Since radar signals vary over time, the model is augmented with multi-layer gated recurrent unit (GRU) cells \cite{turkoglu2021gating} to deal with gradient vanishing arising from signals disappearing with increasing time steps. Moreover, in order to maintain the relationship between the entire time steps, the authors add an attention block \cite{shen2018disan} to the RNN layer. The authors assumed up to 5 targets and 8 interference sources to be experienced at the same time and considered three types of interference (Chirp Sequence, triangle FMCW, and multiple frequency shift keying). In \cite {Ujan_RFI_recognition_2020}, the authors employ different CNN models to mitigate CTI in satellite-to-ground links. The proposed model is trained to identify the sources of received signals (signal of interest) and to classify the modulation schemes with which these signals are modulated (the model is trained to discriminate between the standard DBB-S2 modulation schemes, namely, QPSK, 8-APSK, 16-APSK and 32-APSK). The model was trained with a video stream modulated by a GNU radio, and transmitted using a USRP-N210 \cite{wyglinski2016revolutionizing}. Three different jamming signals -- Continuous Wave Interference (CW), Multi Continuous Interference (MC), and Chirp Interference (C)-- were used to generate interference:
\begin{align}
    CW & = e^{j2\pi f_it} \\ \nonumber
    MC  &  = e^{j2\pi f_{j}t} + e^{j2\pi f_{k}t} \\ \nonumber
    C & = e^{j2\pi \left (\frac{f_l-f_m}{2T}t^2 + f_mt\right)}
\end{align}
where $f_i \dots f_m$ are sweeping frequencies and $T$ the sweeping period, respectively. Fig.~\ref{fig:scalogram_rfi} displays different scalogram plots illustrating the characterisation of clean and noisy signal of interest.

\begin{table*}[h!]
\centering
\begin{tabular}{ |p{3.2cm}|p{3.8cm}|p{3.8cm}|p{3.8cm}|  }
 \hline
 \multicolumn{4}{|c|}{CTI Avoidance} \\
 \hline
 Research Papers & Research aim & Avoidance method  & Technical features\\
 \hline
 \cite{Oyedare_Interference_suppression_2022, hermans2013sonic, IyerDetectingAvoiding} &Interference detection and mitigation & Apply supervised and unsupervised Deep Learning to classify interference patterns & Exploitation of knowledge of modulation schemes and bit corruption patterns.\\ 
 \hline
 \cite{Ristea_fCN_radar_rfi_2020} &Interference classification and mitigation & A fully convolutional neural network to discriminate between different modulation schemes   & Produce clean radar signal from corrupted signal.\\
 \hline
 \cite{Mun_automotive_radar_interfere_2020} &Interference mitigation for radar signals &  Use of recurrent neutral networks (RNN) to recover corrupted Frequency Modulated Continuous Wave (FMCW) radar signals & Increase the capacity of detecting radar targets in the presence of strong ambient noise (Interference) \\ 
 \hline
 \cite{tytgat2014analysis} & Performance improvement in the presence of strong CTI & Multichannel protocol for supporting time-slotted and frequency hopping medium access & Analysis and experimental verification of frequency-based interference avoidance mechanism\\ 
 \hline
 \cite{yung_QF-MAc,KruegerAvoidInterfer}& Overcoming the limitations of TSCH and CSMA/CA-based schedulers &Adaptive channel hopping based on single-hop neighborhood coordination in multi-radio communication & Minimise the effect of interference and maximise network performance in the presence of concurrent networks\\
  \hline
\end{tabular}
\caption{Review of CTI avoidance mechanisms.}
\label{table:1}
\end{table*}

\section{Coexistence}
\label{sec:coexist-ISM}

When multiple wireless technologies concurrently operate in the same frequency band without significantly affecting their performance or the performance of the other technologies, this condition is known as ``coexistence''. Garroppo et al. \cite{Garroppo_expermentCoexistence} carried out a controlled experiment to investigate the extent to which IEEE 802.11 b/g/n, IEEE 802.15.1, and IEEE 802.15.4 specifications enable coexistence. 

\begin{figure}[ht!]
	\centering
	\includegraphics[width=0.50\textwidth]{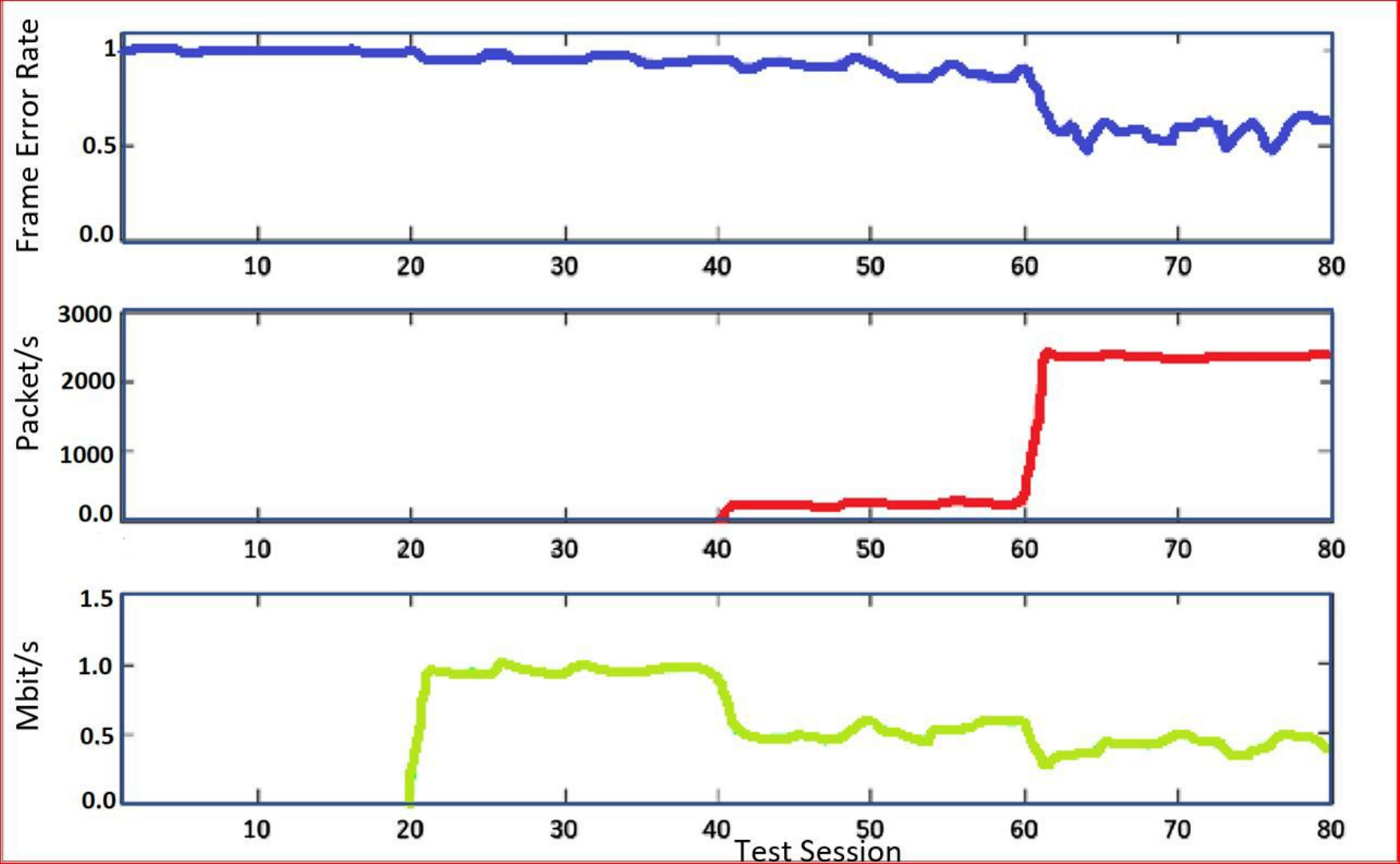}
	\caption{Patterns of interference when three technologies operate concurrently. Bluetooth (bottom), WiFi (Centre), and  IEEE 802.15.4 (top). FIR stands for Frame Error Rate. 1-FER = 1 is equivalent to no frame error rate; 1-FER= 0.5, 50\% frame error rate; 1-FER = 0, 100\% frame error rate .}
	\label{fig:coexistence-test}
\end{figure}

Accordingly, the WiFi devices operate on Channel 6; the Zigbee, on Channel 18; and the Bluetooth, spanning the 2.4 GHz band. As mentioned in Section~\ref{sec:cti}, the three most frequently used non-overlapping WiFi channels are 1, 6, and 11. In the 802.15.4 specification, Channel 18 is totally overlapped with the 802.11 b/g Channel 6. In the beginning, the WiFi and the Bluetooth devices were inactive while two ZigBee devices\footnote{When the context is clear we refer to networks the nodes of which rely on IEEE 802.15.4 compatible radios as low-power networks.} exchanged packets undisturbed. During this time almost all the packets were delivered successfully. After the 20th session, the Bluetooth devices become active and started to exchange packets at 1 Mbps rate. This time the ZigBee devices started to experience moderate packet loss. After the 40th session, the WiFi devices became active, thus causing the other devices to experience appreciable packet losses. The authors gradually increased the WiFi devices' packet transmission rate and by the time they reached 2500 packets per second, the performances of the ZigBee and the Bluetooth devices were severely constrained, the devices most affected were the Bluetooth devices, as can be seen in Fig.\ref{fig:coexistence-test}. Moreover, the experiment results suggest that Bluetooth and ZigBee coexist without appreciably affecting each other's performance. 

Advanced coexistence strategies closely examine opportunities at the physical layer to enable concurrent transmission. For outdoor deployments, the two most important technologies which require coexistence are those based on the IEEE 802.15.4 and IEEE 802.11 b/g/n standards. Medium access in these technologies is based on a clear channel assessment (CCA) which, theoretically, should enable these technologies to share a common medium fairly. However, due to a significant disparity in their transmission power and radio sensitivity, the latter often fail to sense the activities of the former (their default threshold power is much higher than the nominal transmission power of 802.15.4 devices). Even if low-power activities were possible to detect, medium sharing on the basis of CCA alone is not advantageous to WiFi networks due to a big disparity in data rate. Low-power networks achieve much smaller bit rates and significantly longer transmission time, so that their medium occupation time to transmit a single packet will be deemed unfair.

One of the most closely investigated features for enabling coexistence is spectrum reuse. Specifically, WiFi technologies employ OFDM due to its several advantages, including efficient use of a spectrum, resilience to frequency selective fading, and computation efficiency. In OFDM, a channel is divided into multiple orthogonal subcarriers, some of which are used to modulate data and some, as pilot tones. The latter are used to convey a predefined data sequence which is used to determine the difference, or error, between an ideal signal and a received signal. In other words, pilot tones are used for synchronisation and easy signal detection. Each of the data subcarriers can be modulated using BPSK, QPSK, 16-QAM or 64-QAM. One interesting aspects is that, it is possible to selectively avoid using some of these data subcarriers, so that the spectrum occupied by them can be available for the low-power networks to concurrently transmit packets. Intelligent channel assignment is required on both sides. WiFi devices need to determine the subcarriers overlapping with some of the IEEE 802.15.4 channels in order to free them, whereas the low-power devices need to scan the available spectrum to determine the best channels for concurrent transmission.

\begin{figure}[ht!]
	\centering
	\includegraphics[width=0.50\textwidth]{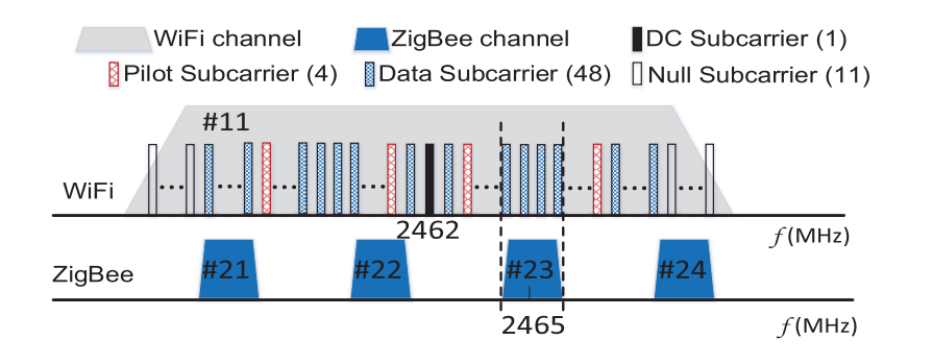}
	\caption{Overlapping OFDM data subcarriers which can be  freed for concurrent transmission by IEEE 802.15.4 low-power networks \cite{guo2020zigfi}.}
	\label{fig:ctc}
\end{figure}

More recently, researchers have started to explore mechanisms which enable direct communications between heterogeneous networks in order to coordinate channel assignments. The approaches are collectively known as cross-technology communication (CTC) \cite{kim2015freebee, chen2020reliable}. Here as well, knowledge of frame structure, medium access mechanism, modulation, channel state information, etc. is exploited to enable direct communication. The prevailing idea is the following: A WiFi device having ample resources senses the presence of a low-power device in the vicinity and encodes a hint as regards its channel utilisation and transmission strategy in its outgoing packets in such a way that the low-power device is able to decode the hint to either adapt its transmission timing or select a complementary channel to avoid CTI \cite{kim2015freebee}.  

In \cite{chen2020reliable}, at a WiFi transmitter a message is deliberately generated, so that when it undergoes the entire modulation and frame structuring, it can be received, demodulated, and decoded as a legitimate IEEE 802.15.4 packet. Accordingly, the message is first  encoded into a set of Quadrature Amplitude Modulation (QAM) constellation points. Secondly, the points are modulated into 48 data sub-carriers using OFDM. Thirdly,  the modulated subcarriers are combined using the Inverse Fast Fourier Transform (IFFT). Fourthly, the time-domain OFDM symbol is prefixed by the cyclic prefix (CP), which is a necessary part of the OFDM symbol to offset the multi-path channel effect. Having thus generating a sequence of WiFi symbols, a packet is transmitted by the WiFi RF radio. On the receiver side, the WiFi header, preamble, and trailer will be regarded as noise and, therefore, will be ignored; the payload, however, will be received, demodulated, and decoded. The process is certainly error prone, but the approach relies on the receiver's (IEEE 802.15.4 ) capability to detect and decode even corrupted symbols.

\begin{figure}[ht!]
	\centering
	\includegraphics[width=0.5\textwidth]{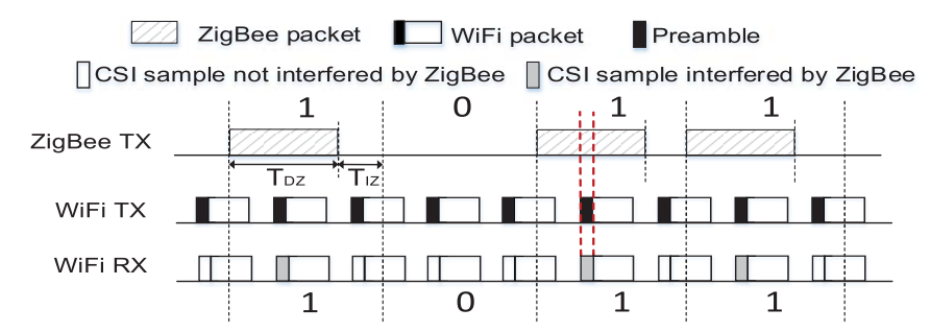}
	\caption{Superimposing IEEE 802.15.4 packets on WiFi packets to create a desirable CSI in order to implicitly enable a direct cross-technology communication \cite{guo2020zigfi}.}
	\label{fig:ctc2}
\end{figure}

In \cite{guo2020zigfi}, the authors exploit Channel State Information (CSI) to enable a low-power device to transmit a coordination message to a resource-rich WiFi device. CSI enables a WiFi receiver to measure the channel status for each OFDM subcarrier during packet reception. Measurement includes phase shift and amplitude variations of a received signal from a reference signal (RSSI is sampled at ca. 31 KHz and phase shift detection is made at 4 MHz \cite{li2017webee}). For the low-power device to successfully transmit a coordination message, it has to overlap its packets with the packet of a nearby WiFi transmitter, so that the superposition of the two signals at the WiFi receiver produces the desired CSI. The idea is illustrated in Fig.~\ref{fig:ctc2}. In the figure, the low-power device wishes to transmit the bit stream 1101 and estimates the transmission pattern of the nearby WiFi transmitter with whose packet it overlaps its own packets, destining them to a common receiver. When the overlap is successful, this will be encoded by the receiver as ``1'', otherwise, as ``0''. In the meantime, the WiFi device will also successfully intercept the packet from its peer, taking into  account the channel state information. In \cite{chen2020reliable}, the authors combine CTC with forward error correction to achieve reliable data dissemination from a WiFi Access Point (AP) to a ZigBee network in the presence of a considerable packet loss. The proposed solution explores chip-level error patterns and attempts to correct  emulation errors by assuming that some errors are more likely to occur than others. The authors claim that chips in some positions of a symbol are more prone to error than others. Accordingly, the authors leverage the cyclic-shift feature of Pseudo Random (PN) sequence and combine parts of correct chips in different sequences into a complete and correct symbol without even accessing the chip information hidden by the hardware.

\begin{figure}[t!]
	\centering
	\includegraphics[width=0.5\textwidth]{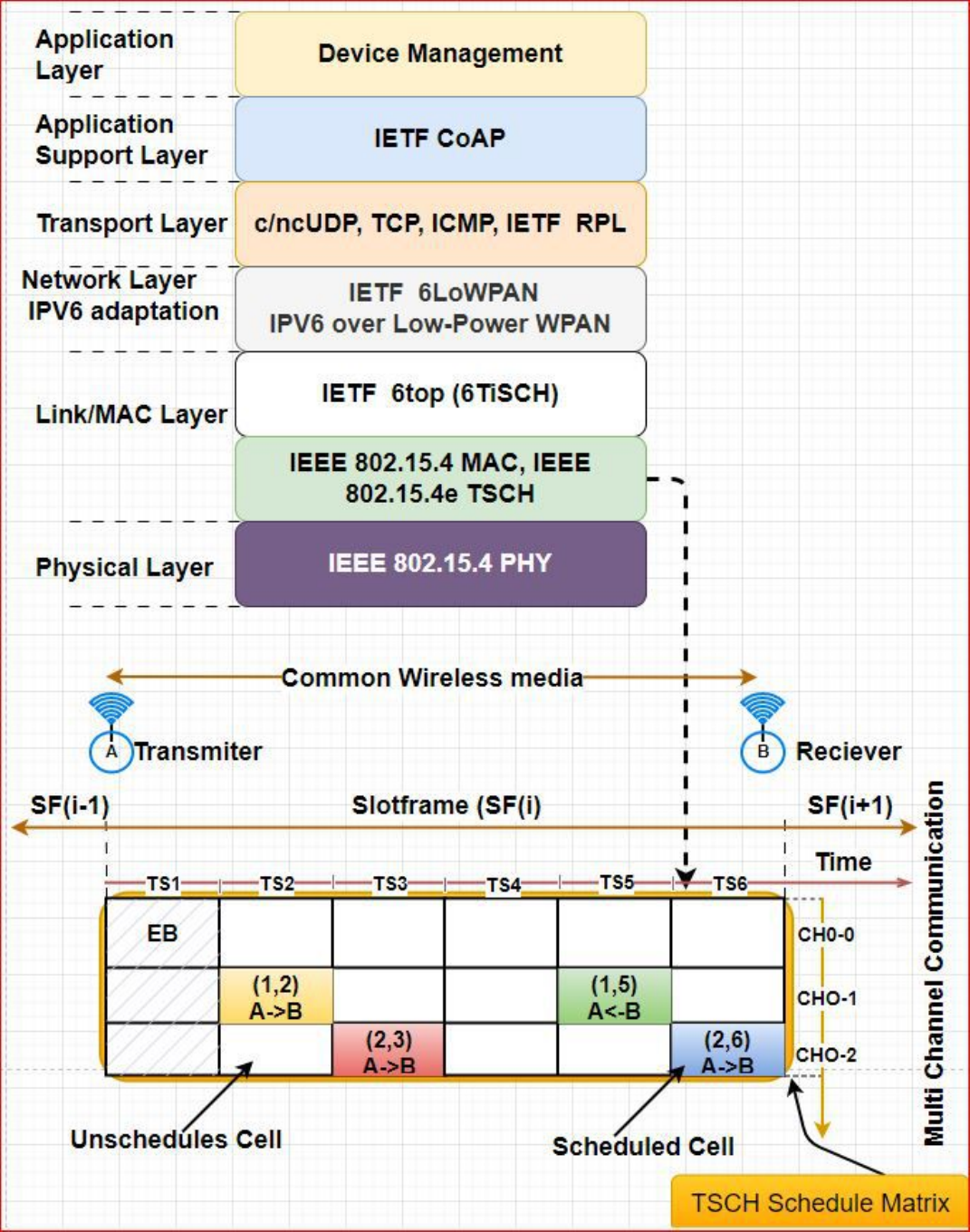}
	\caption{}{Two IoT nodes communicating over the 6TiSCH protocol stack.\footnotemark}
	\label{fig:6tisch_protocol}
  \end{figure}
  \footnotetext{
    *c/cn UDP: Compressed /not compressed user data gram protocol \\
    *CoAP: Constrained Application protocol \\
    *ICMP: Internet Control Massage Protocol \\
    *TCP: Transmission Control Protocol, \\
    *Ts: Time Slot , SF : Slot Frame, \\
    *CHO : Channel offset, EB: Enhanced Beacon, Cell (CHO,TS)   
    }
    
\section{System Support}
\label{sec:system}

The problem of CTI is well-known amongst system software and protocol developers for low-power networks \cite{silva2019operating}. The two widely used runtime environments, namely, CONTIKI \cite{oikonomou2022contiki} and RIOT \cite{baccelli2013riot}, provide system support (message propagation, distributed time synchronisation, beacon propagation, scheduling) for easy implementation and configuration of CTI-aware protocols. The 6LoWPAN \cite{mulligan20076lowpan} protocol stack (Fig. \ref{fig:6tisch_protocol}) likewise accommodates protocols which aim to mitigate CTI. In the next subsection, we concisely present one of the most widely employed medium access protocols, which was initially proposed to deal with CTI in industrial IoT \cite{sisinni2018industrial}.

\subsection{Time-Slotted Channel Hopping}
\label{sec:system_tsch}

Time-Slotted Channel Hopping (TSCH) is one of the most widely used medium access protocols in low-power wireless networks. Initially proposed to mitigate multi-path fading and interference due to electromagnetic noise \cite{ieestd808154e} in industries, it has now become a part of the IEEE 802.15.4 specification and implementations exist for the CONTIKI and RIOI platforms. As its name suggests, TSCH combines the two well-known multiple access techniques, namely, Time Division Multiple Access (TDMA) and Frequency Division Multiple Access (FDMA). The protocol presupposes time synchronisation but gives allowance to some degree of synchronisation error, which is inevitable in resource-constrained devices. 

\begin{figure}[t!]
	\centering
	\includegraphics[width=0.5\textwidth]{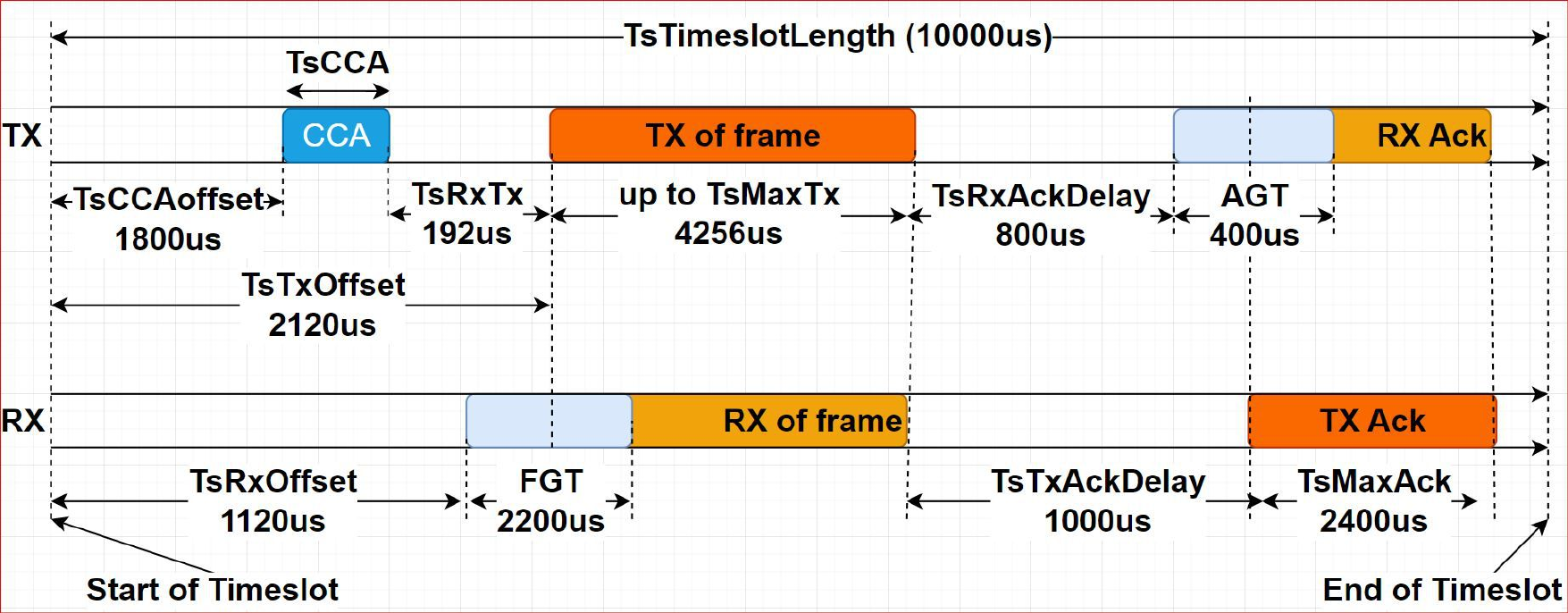}
	\caption{}{Packet transmission between two IoT nodes using the TSCH protocol.\footnotemark}
	\label{fig:tsch_timeslot}
\end{figure}

 
Packet transmission in TSCH takes place in fixed time slots and time slots are organised into frames. The length of a slot is not fixed, as it depends on many factors (including the size of the network) \cite{ieestd808154e}, nevertheless, in the literature a slot length is typically between 10 and  20 ms \cite{Stefano_wsn_tsch}, 10 ms being the most widely adopted. An illustration of a unicast communication between a transmitter and a receiver pair (RX $\leftrightarrow$ TX) (assuming a slot length of 10 ms) is depicted in Fig. \ref{fig:tsch_timeslot}. Referring to the table at the bottom in Fig. \ref{fig:6tisch_protocol}, the rows signify  the channel offsets, whereas the columns, the slot offsets. Similarly, a cell in the table describes a specific medium access configuration or schedule. 

\subsubsection{Packet Reception}
\label{sec:system_packet}

Initially, a receiving node is either in a sleep or an idle state. Packet reception begins when it turns on the radio and prepares for receiving a packet. The transition (TsRxOffset) consists of determining that the medium is free. If it is, the receiver listens for a duration of Frame Guard Time (FGT), waiting for to a valid frame to arrive. If the channel's state does not change after this time,  (i.e., the radio does not detect a signal power exceeding a threshold value), it transits back to an idle or a sleep state (depending on the dynamic power management policy); otherwise, it begins receiving a packet. Following a successful reception, it transits to a transmission state. The transition takes some time (TsTxAckDelay) as the receiver switches the radio from a receiving to a transmission mode and prepares an acknowledgement packet. 

 \footnotetext{
    *CCA: Clear Channel Assessment,\\ 
    *Rx: Receiver,\\ *Tx: Transmitter, \\ *Ts: Timeslot, \\ *TsCCA: Duration of CCA  \\
    *TsCCAOffset: Time to start the CCA operation,\\ *TsRxTx: Transition time to frame transmission, \\ *TsRxAckDelay: Transition time to ACK reception,\\ *TsTxAckDelay: Transition time to ACK transmit, \\ *TxMaxAck: Maximum duration of ACK, \\ 
    *AGT : ACK Guard Time , \\ *FGT: Frame Guard Time   
    }

\subsubsection{Packet Transmission}

A transmitter transits from a sleep or an idle state to a transmit state. The transition consists of making the packet ready for transmission (adding link layer headers to the data frame, performing encryption if security is required, copying the data frame to the radio buffer, and performing clear channel assessment). The time allocated for the transition (TsTxOffset) must be long enough to make a packet ready for transmission, including the time needed for queuing and for the transmission of a preamble and the Start of Frame Delimiter (SFD). Similarly, the transmission duration of the frame that follows the SFD must be equal or smaller than the duration required to transmit the longest frame. The maximum frame length in IEEE 802.15.4 is 128 bytes. Once a packet transmission is over, the node transits to a receive state to await an acknowledgement. The time allocated for the transition is TsRxAckDelay. A minimum amount of time (AGT) must expire to determine whether a packet transmission was a success or a failure. 
 
TSCH deals with interference by relying on channel hopping, using channel offset schedules \cite{yung_QF-MAc}. However, the offset schedule is pseudo-random in that a predefined and globally shared sequence of channels is employed to determine the jump sequence between channels. Equation (\ref{eq:function_mapping}) describes the jumping function: 
\begin{equation}
\label{eq:function_mapping}
C_{ch} = C_{map}[ASN+CH_{off}]mod \hspace{0.1cm}N_{ch}.
\end{equation}
where $C_{ch}$ is the new mapped channel, $ASN$ stands for Absolute slot number, CH$_{off}$ is the channel offset of the communication link between two participant nodes, N$_{ch}$ is the length of the channel hopping sequence; and $C_{map}$ is the channel mapping function. TSCH performs well when the network size is relatively small (relative to the available channels) and the interference magnitude is modest. It is, however, vulnerable to selective external interference such as jamming attacks, as it lacks the capacity to (1) learn the characteristics of the perceived interference and (2) adapt its hopping pattern in accordance with the channel condition. Secondly, the number of active transmission links in the network must be at most N$_{ch}$ to avoid packet collision arising with thin the same network.

\subsection{TSCH Variants}

Channel hopping in multi-channel multi-radio wireless mesh network has been studied in \cite{Tan-MCMR} using spectrum efficiency gain as a performance metric. As can be observed in Fig. \ref{fig:6tisch_protocol}, each packet transmission (and retransmission), takes place in different channel. The channel is selected randomly based on a pseudo-random pattern. The study suggests that the approach reduces interference and improves reliability. Nevertheless, the approach performs well when the interference arises from within the same network. A similar study is carried out in \cite{JavanAdaptiveChanel} using an adaptive frequency channel hopping based on a Model-free and a Model-based schemes. The former assumes that statistics pertaining to channel dynamics are available at design time, whereas the latter typically employs machine learning to establish and react to channel dynamics \cite{Aoudiamodelfree}. Hence, each node in a network is regarded as a learning agent which gradually estimates channel dynamics based on its short-term experience. As a result, the node inclines to favour those channels with higher success rate for its future transmission \cite{JavanAdaptiveChanel}.

Javan et al. \cite{JavanAdaptiveChanel} model CTI as a non-stationary random variable and the task of identifying the next best transmission channel, as a Dynamic Multi-Armed Bernoulli Bandit (Dynamic MABB) process \cite{gupta2011thompson}. The authors then propose an online learning algorithm with tracking ability for computing the best adaptive hopping policy. The work in  \cite{KarolinyHypothesisInterference} attempts to detect and model multiple sources of periodic interference in time-slotted medium access protocols, with the end goal being estimating both the channels and the length of the time windows of future transmissions. The task is formulated as a Multi Hypothesis Tracking problem (MHT) \cite{blackman2004multiple}.

In \cite{gomes_MABO-TSCH}, the authors integrate channel blacklisting with TSCH to mitigate CTI. The proposed approach is decentralised and enables any pair of receiver/transmitter nodes to negotiate a local blacklist, based on the estimation of packet delivery ratio. The channel quality estimation itself is modelled as a multiarmed bandit problem. Dynamic channel selection is carried out by combining three key algorithms: The first algorithm (which is centrally executed and its results are used by a path computation element) is responsible for computing and disseminating the TSCH schedule (channel offsets). This is the basis for all subsequent dynamic channel selections. The second algorithm enables nodes to exchange information about blacklisted channels by piggybacking blacklisting information into data and ACK frames. The third algorithm identifies channels suffering from CTI and blacklists them and maintains the list containing blacklisted channels.

\section{Impacts of Interference}
\label{sec:impacts}

CTI is disruptive. It corrupts packets and inhibits medium access. If packet retransmission is not required, the cost of CTI will be manifested mainly in terms of poor packet reception ratio and high latency. If retransmission is required, the cost will be manifested mainly in terms of high energy cost, short network lifetime, and latency, among others. In order to estimate the energy cost of retransmission, one has to first estimate the transmission and reception cost of a single packet. Nominal values can be obtained from the radio data-sheet once the type of the radio is known, but often this alone is not sufficient because the transmission of a packet involves multiple layers, the management of which varies from operating system to operating system. In the following subsections, we present some experimental studies which highlight the cost of CTI in terms of energy and loss of performance. 

\begin{figure*}[t!]
	\centering
	\includegraphics[width=0.8\textwidth]{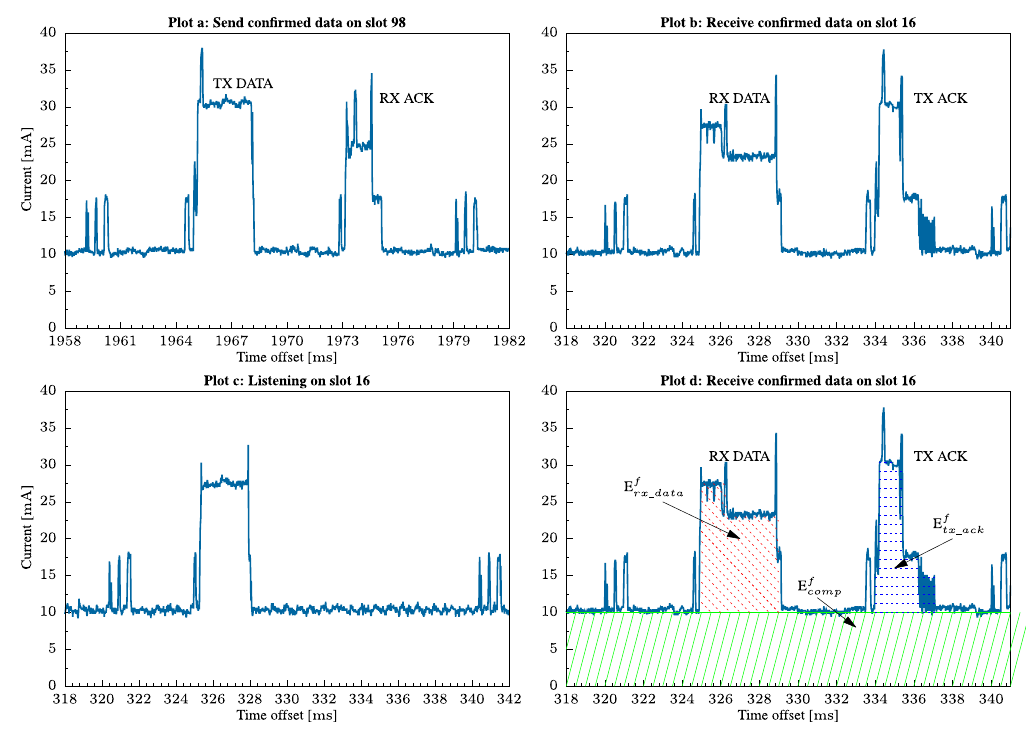}
	\caption{Power consumption for the different TSCH cells, (a) A slot wherein a data packet is transmitted and an ACK packet is received. (b) A slot wherein a data packet is received and an ACK packet is transmitted. (c) An idle time slot. (d) Dissection of overall consumption into the cost of different activities \cite{stefano_TschPowerConsumption}.} 
	\label{fig:power_consume2}
\end{figure*}

\subsection{Energy}

Stefano et al. \cite{stefano_TschPowerConsumption} investigate the energy cost of various activities using OpenMote B platforms \cite{vilajosana2015openmote} and the OpenWSN runtime environment \cite{watteyne2012openwsn}. The runtime environment includes a complete implementation of the 6TiSCH protocol stack (refer Figure \ref{fig:6tisch_protocol}), including TSCH as its medium access protocol. (A similar investigation is carried out using the CONTIKI operating system somewhere else \cite{Sordi_energy}). The OpenMote B platform is based on the CC2538 system-on-chip, which, in turn, integrates an IEEE 802.15.4-compliant 2.4 GHz radio. To measure the current flowing into various sub-components, the authors employed an oscilloscope and disabled all LEDs to exclude their power consumption from further consideration. Thus, the power consumption of a node was segmented into a transmission cost, a receiving cost, a listening cost, and a computation (idle) cost. A transmission cost includes the cost of transmitting a data packet and receiving an acknowledgement packet. Similarly, a receiving cost includes the cost of receiving a data packet and transmitting an acknowledgement packet. As can be seen in \autoref{fig:power_consume2}, tracing the power consumption (current drain) of a single node enables to estimate the beginning and end of a TSCH time slot -- for the experiment, a time slot had a duration of 20 ms -- by considering the receiving and transmitting characteristics of the node. But more importantly, the study clearly shows that the transmission and receiving cost by far dominate the power consumption of all other activities, suggesting that upon experiencing a collusion due to CTI, the cost of packet retransmission is high.

\subsection{Network Performance}

Packet loss and packet delivery latency are two of the most important performance losses resulting from CTI. The latter consists of extended medium access time, extended routing time (packet reaches its destination via multiple links), retransmission delay, and delay due to congestion.

\begin{figure}[ht!]
	\centering
	\includegraphics[width=0.45\textwidth]{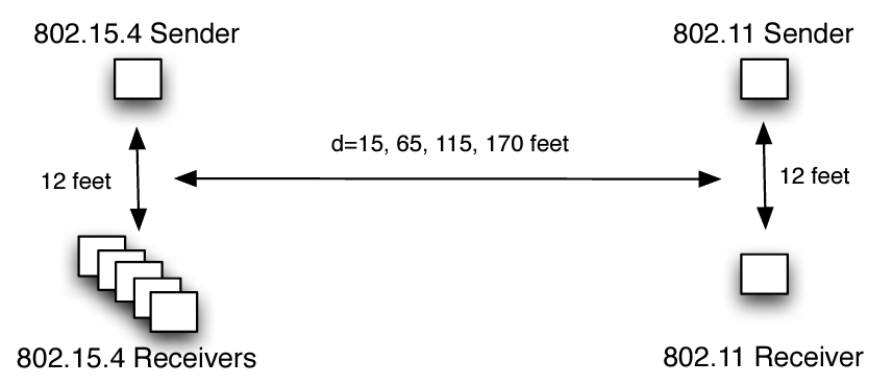}
	\caption{Experimental setup of two heterogeneous networks operating in a self-induced CTI \cite{Liang_surviving_WiFi_interference}.} 
	\label{fig:cti_exp}
\end{figure}

Liang et al. \cite{Liang_surviving_WiFi_interference}  experimentally investigated the impact of CTI on the performance (packet loss and latency) of IEEE 802.15.4 networks when operating in the vicinity of 801.11b and 802.11g networks. A series of experiments were conducted in a quasi-isolated environment (the basement of a big building), thus ensuring that the predominant CTI was from within the experiment setup. The major contribution of this work is the employment of a special narrow band radio (RFMD ML2724) which enabled the interception of RF transmission originating from IEEE 802.15.4 transmitters. The radio enables the bit-by-bit decoding of the RF signal to determine the exact location and patterns of bit errors resulting from CTI. In other words, when placed near an 802.15.4 receiver, the narrow band radio enables to experience the impact of the CTI which the low-power receiver is subjected to. The radio can be tuned to a central frequency between 2400 and 2485 MHz and generates an analog voltage directly proportional to an RF signal energy with a 2 MHz bandwidth. For their experiments, the authors chose Channel 22 for the IEEE 802.15.4 network and Channel 11 for the 802.11b/g networks. A low-power transmitter broadcast packets having 128 byte size to multiple receivers at 75 ms interval. The narrow band radio as well as the low-power receivers kept record of lost packets, packets with CRC errors, and packets which were received successfully. Based on the analysis of these statistics, the authors made the following observation:

\begin{itemize}
    \item Packets originating from  802.11b networks had a much higher impact on the overall 802.15.4 packet reception rate than those originating from IEEE 802.11g networks -- the authors attributed this to the higher transmission rate (i.e., lowers channel occupation time) of 802.11g networks.
    \item The number of lost packets was larger than the number of packets received with bit errors. This suggests that the synchronisation header (SHD) -- a combination of the preamble and Start-of-Frame Delimiter (SFD) -- was more vulnerable to 802.11 interference. This is particularly the case when the cause of the CTI was the IEEE 802.11b network.
    \item Packet transmission latency increased by as much as 13\% to 40\% in the presence of 802.11g and 802.11b traffic, respectively.
    \item By contrast, the WiFi networks suffer a modest amount of packet loss when the transmission power of the IEEE 802.15.4 transmitter was at its highest and its distance to the 802.11 network was approximately 15 feet.
\end{itemize}

Hithnawi \cite{hithnawi2014understanding} carried out a series of experiments similar to that of Liang et al., but arrived at a slightly different conclusion. For their experiments, the authors chose an anechoic chamber having dimensions: 7 m $\times$  4 m $\times$ 4 m. An IEEE 802.15.4 transmitter transmits variable sized packets (20, 40, 100 bytes) at variable transmission power levels (0 dBm, -3 dBm, and -10 dBm) and intervals. At the same time a nearby WiFi device was interacting with a router in different configurations: accessing the medium with and without a clear channel assessment; and transmitting TCP and UDP packets at different intervals and with a transmission power which was gradually adjusted from -20 dBm to 20 dBm. The authors observed that configuring the WiFi network to perform with or without clear channel assessment had little impact on the reduction of CTI or its impact, as the networks (802.11 b/g) failed to detect the activities of the nearby low-power networks. On the other hand, the traffic pattern and size of the WiFi networks had a considerable impact. Thus, when their traffic was modest, it made no appreciable impact on the performance of the low-power networks; but when the traffic was dense and frequent (packets transmitted every 7 ms), it reduced the performance of the low-power networks by up to 20\%.    

\section{Conclusion}
\label{sec:conclusion}

The joint deployment of heterogeneous wireless networks is presenting a great opportunity to support a wide range of critical applications. In industrial IoT, wireless sensors, mobile robots, and other objects can seamlessly interact to facilitate safe and efficient operations. In water quality monitoring, Unmanned Aerial Vehicles, Unmanned Surface Vehicles, and wireless sensor networks can be deployed to monitor the quality of an extensive water body. Because of the safe operation of the mobile robots, the UAVs, and the USVs is critical not only to fulfill the purpose for which they are deployed but also because of their cost and the damage they may cause in case of error in their navigation plan, often they require highly reliable links. For these reasons, they rely on wireless links that afford them with high transmission power and wide bandwidth. By contrast, most existing sensing networks are low-power networks. Consequently, when such heterogeneous networks operate in close proximity to each other sharing a common spectrum, a cross-technology interference ensues and the impact of this interference is not reciprocal. 

In this paper we reviewed various strategies to detect and deal with cross-technology interference. The detection strategies range from high-level strategies which attempt to statistically model network traffic and estimate traffic patterns and interval to enable packet scheduling to low-level signal processing which attempt to take advantage of knowledge of modulation and channel coding to analyse patterns of symbol corruption and bit errors. Similarly, the coexistence strategies strive to enable concurrent operations by employed high-level as well as low-level strategies. More recently, researchers have also started investigating ways for enabling direct communication between heterogeneous networks to coordinate channel assignment and mitigate cross-technology interference.

In this paper we have not investigated the cost of managing cross-technology interference. This is rather important considering the fact that some of the strategies are low-level and require a considerable computation. For example, some of the strategies aiming to enable direct cross-technology communication, rely on knowledge of the communication pattern of active networks to overlap packets and, thereby, convey a message on signals superimposed by the heterogeneous networks. This requires precise timing and fine-grained estimation. Our plan for the future is to address such issues and to quantitatively compare the performance of some of the proposed approaches.

\section*{Conflict of Interest}
 On behalf of all authors, the corresponding author states that there is no conflict of interest.
 
\balance
\bibliography{library}
\end{document}